\documentstyle[axodraw,11pt,amssymb]{article}
\newcommand{\sect}[1]{\setcounter{equation}{0}\section{#1}}

\newfont{\got}{cmfrak at 11pt}

\newfont{\lil}{rsfs10}

\newfont{\sss}{cmu10 at 11pt}
\newfont{\sic}{cmu10 at 14pt}
\newfont{\som}{cmu10 at 20pt}
\newfont{\bfm}{cmbxti10 at 11pt}
\newfont{\cmna}{cmdunh10 at 14pt}
\newfont{\cmn}{cmdunh10 at 11pt}
\newfont{\san}{cmssq8 at 11pt}
\newfont{\bon}{cmvtt10 at 11pt}
\def\rf#1{(\ref{eq:#1})}
\def\lab#1{\label{eq:#1}}
\def\nonu{\nonumber}
\def\br{\begin{eqnarray}}
\def\er{\end{eqnarray}}
\def\be{\begin{equation}}
\def\ee{\end{equation}}

\def\lb{\lbrack}
\def\rb{\rbrack}

\def\({\left(}
\def\){\right)}

\newcommand{\ct}[1]{\cite{#1}}
\newcommand{\bi}[1]{\bibitem{#1}}

\relax

\def\Tr{\mathop{\rm Tr}}

\newcommand{\sbr}[2]{\left\lbrack\,{#1}\, ,\,{#2}\,\right\rbrack}

\newcommand\bra[1]{\langle \, {#1}\, \mid}
\newcommand\ket[1]{\mid \, {#1} \, \rangle}

\def\a{\alpha}

\def\b{\beta}

\def\ba{{\bar a}}

\def\bz{{\bar z}}

\def\d{\delta}

\def\g{\gamma}

\def\h{{1\over 2}}
\def\l{\lambda}
\def\L{\Lambda}

\def\o{\over}

\def\pa{\partial}
\def\pr{\prime}

\def\ra{\rightarrow}

\def\tp0{\Theta_{+}^{(0)}}
\def\tm0{\Theta_{-}^{(0)}}

\def\u2{\mid u\mid^2}

\def\vp{\varphi}

\def\cgh{{\hat {\cal G}}}

\def\cv{{\cal V}}

\def\f#1#2#3 {f^{#1#2}_{#3}}

\def\win1{{\sf w_{1+\infty}}}

\def\Win1{{\sf W_{1+\infty}}}

\def\IZ{{\mathbb Z}}

\def\one{\hbox{{1}\kern-.25em\hbox{l}}}
\def\0#1{\relax\ifmmode\mathaccent''7017{#1}%
B        \else\accent23#1\relax\fi}

\def\PLB#1#2#3{{\sc Phys. Lett.} {\bf #1B} (#2) #3}
\def\JMP#1#2#3{{\sc J. Math. Phys.} {\bf #1} (#2) #3}

\def\dw{\partial_w}

\def\dbw{\partial_{\bar w}}

\def\sinh{{\rm sinh}}

\def\Tr{{\rm Tr}}
\begin{document}
\begin{titlepage}
\vspace*{-1cm}

\noindent
August, 2002 \hfill{IFT-P.056/02} 

\hfill{SISSA/59/02/EP}

\hfill{hep-th/0208175}  

\vskip 1.5cm

\vspace{.2in}
\begin{center}
{\large\bf Construction of exact Riemannian instanton solutions}
\end{center}

\vspace{.5cm}

\begin{center}
L. Bonora~$^1$, C. P. Constantinidis~$^2$, L. A. Ferreira~$^3$ and
E. E. Leite~$^3$

\vspace{.5 in}
\small

\par \vskip .2in \noindent
$^{(1)}$~International School for Advanced Studies (SISSA/ISAS)\\
Via Beirut 2-4, 34014, Triste, Italy, \\
and INFN, Sezione di Trieste 

\par \vskip .2in \noindent
$^{(2)}$~Universidade Federal do Esp\'irito Santo - UFES\\
Departamento de F\'isica - CCE\\
Goiabeiras, 29060-900, Vit\'oria - ES, Brazil\

\par \vskip .2in \noindent
$^{(3)}$~Instituto de F\'\i sica Te\'orica - IFT/UNESP\\
Rua Pamplona 145\\
01405-900  S\~ao Paulo-SP, Brazil\\

\normalsize
\end{center}

\vspace{.5in}

\begin{abstract}
We give the exact construction of Riemannian (or stringy) instantons, 
which are classical solutions of $2d$ Yang-Mills theories 
that interpolate between initial and final string configurations.  
They satisfy the Hitchin equations with special boundary
conditions. For the case of $U(2)$ gauge group those equations can 
be written as the sinh-Gordon equation with a delta function source. 
Using techniques of integrable theories based on
the zero curvature conditions, we show that the solution is a condensate of
an infinite number of one-solitons with the same topological charge and
with all possible rapidities.     
\end{abstract}
\end{titlepage}

\section{Introduction}

In this paper we intend to prove the existence of {\it Riemannian}
or {\it stringy instantons}. The name is due to the fact that 
these are classical solutions of a 2D $U(N)$ YM theory that 
interpolate 
between initial and final string configurations. In other words
they describe Riemann surfaces with punctures, where the latter 
represent the asymptotic entering and exiting strings.
In the simplest case ($N=2$), proving the existence of 
Riemannian instantons amounts to finding exact solutions of the 
sinh--Gordon equation with a delta--function source (see below for more
details). To be definite let $a=a(z)$  be a polynomial in the 
complex variable $z$, with distinct roots, and let us introduce a 
new variable $\zeta$ defined by $\frac{\d \zeta}{\d w} = \sqrt  a$, 
where $z= {\rm e}^w$, $w$ being the coordinate on an infinite 
cylinder. The equation we want to solve is
\be 
\pa_\zeta \pa_{\bar \zeta} u - 2g^2 \sinh~ 2u= - \frac{\pi}{4} \delta (a) 
(\pa_\zeta a)(\pa_{\bar \zeta}\bar a) \lab{sinhdelta}\,.
\ee 
which has to be understood in the sense of complex distribution 
theory. In this equation $g$ is a constant coupling.
An equivalent way to state \rf{sinhdelta} is to write the usual 
sinh--Gordon equation
\be
\pa_\zeta \pa_{\bar \zeta} u - 2g^2 \sinh~ 2u= 0\lab{sinhgordon}
\ee
and to look for solutions which, near the zeroes of $a$, behave like
\be
u \sim - \frac{1}{2} \ln |a|, \quad\quad {\rm for} 
\quad\quad a\sim 0
\lab{boundary}\,.
\ee
We will also require that $u$ vanishes as $z\to 0$ and $z\to \infty$.
This kind of equation was met for the first 
time in the framework of Matrix 
String Theory in \ct{wynter,GHV,bbn1,bbn2}, and solutions  to these
equations, satisfying
\rf{boundary}, were shown to exist  only numerically.
In particular in \ct{bbn2} this was done in the framework of a 
square lattice approximation, assuming a very simple form for $a$. 
The same type of solutions appeared in the context of form factors and
correlation functions for the Ising model in \ct{fredholm2}.   

This paper is devoted to proving the existence of solutions to the 
above equations, with the desired boundary conditions, in an analytic
way, and to give their closed expressions in terms of the modified Bessel
function $K_0$. In fact, the validity of the solution relies on some
non-linear differential identites satified by integrals of $K_0$,
which to our knowledge, have not appeared  in the literature. 

The central ideas of the proof is (1) to use the Leznov-Saveliev
approach and (2) to view the solution as a condensate of solitons. 

More in detail, we begin by writing the 
sinh-Gordon equation \rf{sinhdelta} in terms of  zero curvature 
conditions, which include besides the usual Lax-Zakharov-Shabat
equation a second relation leading to non-local conservation
laws. Such generalized zero curvature conditions follow from the
ideas proposed in \ct{afs} to study integrable theories in any
dimension. 
Once we have the equations of motion written in terms of a zero
curvature condition, we  utilize the Leznov-Saveliev
method to construct the corresponding Riemannian instanton solution.
This method uses the fact that the dynamical variables
of the system are contained in the zero curvature potentials:
the sinh-Gordon $\vp$ field appears as a parameter of the
group element we use in order to write the flat connetion $A_\mu$
as
\be
A_\mu = - \pa_\mu WW^{-1} \: = \: \mbox{\rm function of a group
element} \; \g,  \qquad \g \; = \; e^{\vp \, H^0 + \nu C}  .
\ee
Thus, due to the path independence encoded in $F_{\mu \nu}=0$, we
are able to write the group element $W$ in distinct forms. This,
compounded with some properties of the Kac-Mooody algebra, 
leads us to a simple algebraic relation for 
the ``group parameters'' $\vp$ and $\nu$, i.e.
\be
\bra{\l} \g ^{-1} \ket{\l} \;=\; \bra{\l} \g_{+} N_{+} 
M_{-}^{-1} \g_{-} \ket{\l} 
\ee
where the elements $\g_\pm, \, N_+$, and $M_-$ have
nice properties in terms of Kac-Moody algebra representations.
To determine the two parameters $\vp$ and $\nu$,
we make use of two highest weight representations. This
provides us with a relation for $\vp$ in terms of the
expected values of the group elements $N_{+}$ and
$M_{-}$. So, solving the sinh-Gordon equation is
equivalent to furnishing these two elements.
Once this is done, we conveniently choose the parameters  
in such a way that the boundary conditions \rf{boundary} are 
satisfied.  

The key point in this construction is that in order to obtain the 
desired solution we must choose the constant group elements of the
solitonic specialization of the 
Leznov--Saveliev construction as an infinite product of exponentials of vertex 
operators. The product is in fact a continuous one, since it involves all
possible values of the rapidities of the one-solitons. In addition, all
one-solitons entering the expansion have the same topological
charge. This leads us to interpret 
such configuration as a condensate of solitons. As it was realized 
in \ct{fredholm1}, this type of solution 
can be written as  a Fredholm determinant, and our solution is similar 
to the one found in \ct{fredholm2}, where correlation functions 
of the Ising model were obtained in terms of $\tau^{(N)}$ functions 
of the sinh-Gordon model, with $N \rightarrow \infty $. 
In order to arrive at the true solution a continuum limit 
must be taken for the condensate of solitons and the Fredholm 
determinant must be rewritten as an infinite series of integrals, 
whose convergence conditions are studied and particular 
normalizations are fixed in order to satisfy the required boundary 
conditions. The solution heuristically derived in this way is finally
shown to satisfy the equation \rf{sinhdelta} or \rf{sinhgordon}
plus \rf{boundary}.

Before we enter into the very existence proof 
it is worth reviewing the framework where the eq. \rf{sinhdelta}
arises and plays a fundamental role. Matrix String Theory 
(MST) \ct{motl,BS,DVV} is the theory that arises upon 
compactifying Matrix Theory \ct{BFSS} on a circle, \ct{WT}. 
It is expected to be a nonperturbative 
version of type IIA string theory. An attempt to substantiate such
conjecture was started in \ct{wynter,GHV} and completed in
\ct{bbn1,bbn2,bbnt1}, where it was shown that a correspondence
between MST and type IIA theory exists not only at the tree level,
but that actually MST contains the full perturbative expansion of 
type IIA 
string theory. It was in this context that  \rf{sinhdelta} appeared.

Looking for classical solutions of MST that preserve half supersymmetry,
the following system of equations was found
\br
&& F_{w\bar w} -i g^2 [X , \bar X]=0 \nonu\\
&& D_w\bar X=0,\quad\quad D_{\bar w} X=0 \lab{rieminst}
\er
where $F_{w\bar w}$ denotes the curvature of a connection $A_w$ and 
$A_{\bar w}$, while $X$ is an $N\times N$ matrix and $\bar X$ its hermitean 
conjugate. $D_w,D_{\bar w}$ denotes the covariant derivatives with respect to
$A_w,A_{\bar w}$.
 \rf{rieminst} may be called {\it Hitchin equations}, because they were 
discussed first by Hitchin in a different context \ct{hitchin}, or {\it 
Riemannian instanton equations} because of their geometrical interpretation. 
To elucidate
this terminology and the importance of these equations let us consider
the simplest case, in which the gauge group is $U(2)$. The problem
to be solved is finding a couple $(A,X)$ that satisfies
 \rf{rieminst}. To this end we choose the following ansatz
\be 
 X=Y^{-1}MY,\quad\quad A_w=i\dw Y^\dagger (Y^{-1})^\dagger, \lab{ans}
\ee
where $Y$ is a suitable matrix $\in SL(2,{\mathbb C})$, and $M$ is the 
following $2\times 2$ matrix
\be
M=\left(\matrix{0 &a\cr 1 &0\cr}\right)\lab{m22}
\ee
where $a$ is a function on the complex plane.
As a consequence of the equation $D_{\bar w} X=0$, it follows that
$\partial_{\bar w} a =0$, i.e. $a$ is holomorphic in $z$ (at least
for finite $z$). As explained above,
we will assume that $a$ is a polynomial in $z$ with distinct roots. Now, 
given such an
$a$ we want to find $Y$ so that \rf{rieminst} is satisfied.
We parametrize $Y$ as $Y = 
\left(\matrix{ e^p & 0\cr 0&e^{-p}\cr}\right)$ 
where $p= \frac{u}{2} + \frac {1}{4} ln |a|$, and $u$ is a function to be
determined. Then using \rf{ans} we find  
\be
X= \left( \matrix{ 0 &a e^{-2p}\cr e^{2p}&0\cr} \right),\quad\quad
A_w = i \dw p \left(\matrix{1&0\cr 0&-1\cr}\right)\lab{XA}
\ee
Now it is easy to verify that the first equation in \rf{rieminst} 
implies
\be
2\dw \dbw p - g^2 \Big( e^{4p} - |a|^2 e^{-4p}\Big) =0\lab{presinh}
\ee
Inserting the explicit form of $p$ and the change of variable $w \to \zeta$,
s.t. $\frac{\d \zeta}{\d w} = \sqrt a$, one can rewrite \rf{presinh}
as  \rf{sinhdelta}.
If $u$ is a smooth solution of this equation, the couple $(X,A)$ is a 
solution of \rf{rieminst} which is smooth everywhere except perhaps at 
infinity. Now, the important point is that the matrix $M$ represents a 
branched covering of the $z$--plane. This is seen by diagonalizing $M$ by 
means of a matrix
in $SL(2,{\mathbb C})$: 
\be
M = S {\widehat M} S^{-1}, \quad\quad {\widehat M}= \left(\matrix{\sqrt a&0\cr
    0 &-\sqrt a 
\cr}\right),\quad\quad S = \frac{i}{\sqrt 2} \left(\matrix{
a^{\frac{1}{4}}& a^{\frac{1}{4}}\cr
a^{-\frac{1}{4}}&-a^{-\frac{1}{4}}\cr}\right).\lab{SMS}
\ee
The two eigenvalues of $M$ represent the two branches of the equation
\be
y^2= a\lab{hyper}
\ee
which is the defining equation of a hyperelliptic Riemann surface,
with branch points corresponding to the roots of $a$.
Therefore the solution at issue represents a Riemann surface, which 
justifies the adjective in the name {\it Riemannian instanton}. The 
instanton nature of this solution is due to the fact that  
\rf{rieminst} are the two dimensional reduction of the YM 
self--duality equation in 4D.
In \ct{bon} this example was analyzed in great detail. It was shown there
that, if the $u$ solution of \rf{sinhdelta} satisfies the boundary
conditions stated at the beginning of this introduction, 
the matrix $X$,
outside the branch points of $a$ and when $g\to\infty$ , 
becomes $\widehat M$ up to 
a unitary transformation. This fact plays a crucial role
in establishing the correspondence between MST and type IIA theory,
see \ct{bbn1,bbn2}. 

It is therefore of upmost importance to establish the existence of 
the above solutions of  \rf{rieminst} and therefore of
the corresponding $u$ solutions of  \rf{sinhdelta}.
On the other hand, it is clear that showing the existence
of exact solutions of  \rf{sinhdelta} is a problem interesting in 
itself.

The paper is organized as follows. In section 2 we apply the 
Leznov--Saveliev method to our problem and define the soliton
condensate. In section 3 we verify that what has been heuristically 
constructed in section 2 is in fact the looked for 
solution. A few appendices are devoted to clarify some technical 
points utilized in the course of the proof.


\sect{Construction of solution through Leznov-Saveliev algebraic method}

In this section we review the Leznov-Saveliev method \ct{ls} 
for construction of solutions of Affine-Toda type theories, 
based on the zero curvature formulation of two dimensional 
integrable systems. Even though we work in two dimensions we stick to 
the point of view of higher dimensional integrable models, 
which can be constructed from two potentials, 
$A$ and $B$ \ct{afs}. Among other things such an approach  leads to
the construction  
of new conserved currents, not obtained from the usual two dimensional 
formalism \ct{afs}.

\subsection{Zero curvature condition}
\label{zero}

The eq. \rf{sinhdelta} admits  a representation in terms of the
following zero curvature conditions 
\br
F_{\mu\nu}&=& \pa_{\mu} A_{\nu} - \pa_{\nu} A_{\mu} + 
\sbr{A_{\mu}}{A_{\nu}} = 0 \lab{zc1}\\
D^{\mu}B_{\mu} &=&  \pa^{\mu} B_{\mu} + \sbr{A^{\mu}}{B_{\mu}} =
0 \lab{zc2}
\er
In two dimensions the condition \rf{zc1} is the well known
Lax-Zakharov-Shabat equation. In dimensions higher than two the
eqs. \rf{zc1} and \rf{zc2} were shown to be sufficient local
conditions for the vanishing of the generalized zero curvature
equations relevant for higher dimensional integrable theories
\ct{afs}. Here we apply eqs. \rf{zc1}-\rf{zc2}  for the
two dimensional model \rf{sinhdelta}, and show that the relation
\rf{zc2} leads to non-local conservation laws. The procedure applies equally
well  to a wide class of two dimensional integrable models, like the abelian
and non-abelian Toda models (affine or not), 
possessing a representation in terms of Lax-Zakharov-Shabat eq. \rf{zc1}. 
That equation 
can be enriched by the extra conservation laws \rf{zc2} without any
further restriction in their dynamics, and we plan to analyse that in more
details in a future publication.           

Let $\cgh$ be an affine  $sl(2)$ Kac-Moody algebra, with basis 
$H^m, T^m_{\pm}, D, C$, satisfying the commutation relations
\rf{sl2kmcomrel}. We take the local zero curvature potentials as
\br
A_{w} \: \equiv \:  -\pa_{w}\g \g^{-1} + E_{-1} &\qquad&
A_{\bar{w}} \: \equiv \: \g E_{1} \g^{-1} \nonu\\
B_{w} \: \equiv \: P^{\psi}\(E_{-1}\) &\qquad&
B_{\bar{w}} \: \equiv \: P^{\psi}\(\g E_{1} \g^{-1}\)
\lab{ori}
\er
where
\be
\g \equiv e^{\vp H^0 + \nu C}
\lab{gammadef}
\ee
and
\be
E_1 \: \equiv \:   g \; T_{+}^0 + g \, \ba\({\bar z}\) \; T_{-}^1 \qquad \qquad
E_{-1} \: \equiv \:  g \, a\(z\)  \; T_{+}^{-1} + g \, T_{-}^0
\lab{epm1}
\ee
Following \ct{afs} we have taken $A_{\mu}$ and $B_{\mu}$ to belong to
a non-semisimple Lie algebra formed by 
the $sl(2)$ Kac-Moody algebra and an abelian ideal which transforms
under its adjoint representation $P^{\psi}$, i.e. 
\br
\sbr{T_a^m}{T_b^n} &=& f_{ab}^c T^{m+n}_{c} + \Tr\(T_aT_b\) C
\delta_{m+n,0}\nonu\\
\sbr{T_a^m}{P^{\psi}\(T_b^n\)}&=& P^{\psi}\(\sbr{T_a^m}{T_b^n}\)\nonu\\
\sbr{P^{\psi}\(T_a^m\)}{P^{\psi}\(T_b^n\)}&=& 0
\er

With this choice,  
the  zero curvature conditions \rf{zc1}
lead us to the system of differential equations
\br
\pa_{\bar{w}}\pa_{w}  \vp &=& 
g^2\, \(  e^{2\vp} -  \mid a \mid ^2 \; e^{-2\vp} \)
\lab{phieqmot}\\
\pa_{\bar{w}}\pa_{w}  \nu &=&   g^2 \, \mid a \mid ^2 \;  e^{-2\vp} \, .
\lab{eqmotb3}
\er
The identification
\be
\vp = 2 p \equiv  u + \h \ln   \mid a\mid
\lab{phiurel}
\ee
turns \rf{phieqmot} into the equation
\be
2\pa_{w}\pa_{\bar w} p  = g^2
\( e^{4p} - \mid a \mid^2 e^{-4 p} \)  \ee
This is exactly the sinh-Gordon equation \rf{sinhdelta} once we
make the change of variables $w \to \zeta$, such that (see Appendix
\ref{app:delta}) 
\be
\frac{d \zeta}{d w} = \sqrt{a} \qquad \qquad \frac{d {\bar
\zeta}}{d{\bar w}} = \sqrt{\ba} \lab{zeta}\ee

As we point out in the introduction, the sinh-Gordon equation
with source is equivalent to the homogeneous equation
together with the following
boundary conditions
\be
u \sim -\h \ln \, \mid a \mid \qquad \vp \sim {\rm finite}\qquad\qquad
{\rm for} \qquad\qquad
a\sim 0
\lab{boundarycond}
\ee
So, $u$  must diverge logarithmically at the zeroes of $a$, and from
\rf{phiurel} it follows that $\vp$ should be finite there. 
On the other hand, far away from any zero of $a$ we need
\be
u \sim 0 \qquad \qquad \vp \sim \h \ln \, \mid a \mid \qquad\qquad{\rm for}
\qquad \qquad
a \sim \infty
\lab{boundarycond2}
\ee

One can check that the condition \rf{zc2} is trivially satisfied by
the potentials \rf{ori}, i.e. it holds true for any field
configuration including those which are not solutions of the equations
of motion \rf{phieqmot}-\rf{eqmotb3}. However, due to \rf{zc1} it
follows that the connection  $A_{\mu}$  is flat, and so there is
a group element $W$ such that
\be
A_{\mu} \equiv - \pa_{\mu} W W^{-1} \, .
\lab{wdef}
\ee
Consequently, it follows from \rf{zc2} that the currents 
\be
J_{\mu} \equiv W^{-1} B_{\mu} W
\lab{nonlocalcur}
\ee
are conserved
\be
\pa^{\mu} J_{\mu} =0
\ee
The group element $W$ only exists for field configurations that
satisfies the equations of motion, and it is  non-local
in the field variables. Consequently,  the currents  \rf{nonlocalcur}
are non-local. We intend to study the properties of these currents, in
a wide class of models, in a future publication.  

\subsection{Leznov-Saveliev construction}
\label{lezno}

The Leznov-Saveliev method uses some features of the Kac-Moody 
algebra representations, as well as some geometrical properties 
due to the zero curvature condition \rf{zc1}.  
Since $F_{\mu \nu}=0$, it follows that the integration necessary
for obtaining the group element $W$, introduced in \rf{wdef}, is path
independent. That allows us to write this group element
in different ways. We choose two different
paths in order to integrate \rf{wdef}, and get
\be
W = g_1 = \g g_2 \qquad \qquad \qquad \g = g_1 g_2^{-1}
\lab{g1g2def}
\ee
This means that we have two equivalent decomposition
for the potentials $A_\mu$:
\be
\pa_{\mu} g_1 g_1^{-1} \:\: = \:\:
\pa_{\mu} W W^{-1} \:\: = \:\:
 \pa_{\mu} \g \g^{-1}
+ \g \pa_{\mu} g_2 g_2^{-1}\g^{-1} \lab{twow}
\ee

Next we recall some aspects of the $\widehat{sl}(2)$ Kac-Moody
algebra. An  affine Kac-Moody algebra possesses integer gradation such
that \ct{kac,olive0}
 \be \cgh = \cgh_{+} \, \oplus \, \cgh_{0} \, \oplus \, \cgh_{-} \, \ee where
$\cgh_{0,\pm}$ are the eigensubspaces of the grading operator \be
Q\equiv \h H^0 + 2 D \; ; \qquad \qquad \sbr{Q}{\cgh_n} = n \;
\cgh_n \ee with \be \cgh_{+} \, = \,\bigoplus \limits_{n>0} \,
\cgh_n \qquad \qquad \cgh_{-} \, = \,  \bigoplus \limits_{n<0} \,
\cgh_n \ee

Observe that, in terms of the eigensubspaces $\cgh_{0,\pm}$, the
potentials \rf{ori} have the decomposition
\be A_w \: \in \: \cgh_{0} \oplus \cgh_{-}
\qquad A_{\bar{w}} \: \in \: \cgh_{+} \qquad \qquad B_w \: \in \:
P^{\psi}\(\cgh_{-}\) \qquad   
B_{\bar{w}} \: \in \: P^{\psi}\(\cgh{+}\)
\lab{potgrad}\ee

We shall analyse how this potentials acts on a representation
of the algebra.
Consider a representation of the $\widehat{sl}(2)$ Kac-Moody
algebra. Inside it, there is a highest weight state $\ket{\l}$ on
which \be T^n_{+} \, \ket{\l} = 0 \quad n \geq 0 \qquad \qquad
T^n_{-} \, \ket{\l} = 0 \quad n > 0  \, . \lab{accao}\ee
The scalar and spinor representations of the affine 
$\widehat{sl}(2)$ Kac-Moody algebra
have highest weight states $\ket{\l_0}$ and $\ket{\l_1}$ which
satisfy
\br
C \ket{\l_0} = \;  \ket{\l_0} \qquad &&\qquad 
H^0 \ket{\l_0} = 0 
\lab{l0action} \\
C \ket{\l_1} = \; \ket{\l_1} \qquad &&\qquad 
H^0 \ket{\l_1} = \; \ket{\l_1}   
\lab{l1action}
\er

Therefore, using \rf{twow}, \rf{potgrad} and \rf{accao}
we see that the elements $g_1, \, g_2$ were taken in such a
form that
\be
\pa_{\bar{w}} g_1 g_1^{-1} \ket{\l} =0 \qquad \qquad
\bra{\l} \pa_w g_2 g_2 ^{-1} =0
\ee
This means that these group elements evaluated on
a heighest weight representation are holomorfic functions
of the coordinates
\be
g_1^{-1}  \ket{\l} \, \equiv \, {\rm f}\(w \) \qquad \qquad
\bra{\l}  g_2  \, \equiv \, {\rm f} \( \bar{w} \) \,.
\ee

With these ingredients at hand, we are interested in
evaluating the expectation value  
\be
\bra{\l} \g^{-1} \ket{\l} = \bra{\l} g_2 g_1^{-1} \ket{\l}
\lab{pressol}
\ee
that descends from  \rf{g1g2def}. To this end
we perform the Gauss type decomposition
\be
g_1 \equiv N \g_- M_{-} \qquad g_2 \equiv M \g_+ N_{+}
\lab{gauss}
\ee
with
\be
\g_{\pm} \in \exp\( \cgh_0\) \qquad
N \, , \,  N_{+}\in \exp\( \cgh_{+}\) \qquad
M \, , \,  M_{-}\in \exp\( \cgh_{-}\) \, .
\ee
So, using  \rf{wdef}, \rf{twow}, \rf{gauss}
and \rf{accao} we compare the gradings of the 
potentials and of the group elements $g_1, \, g_2$, and
with these we 
determine the action of $\g$ \rf{pressol} 
on the representation states.

Comparing the gradings of the potentials $A_\mu$
with the ones derived from \rf{twow} -- \rf{gauss}, we get
\br
\pa_{\bar{w}} \g_{-} \g_{-}^{-1} &=& 0 \qquad  \qquad
\pa_{\bar{w}} M_{-} M_{-}^{-1}= 0 \qquad  \qquad
\pa_{\bar{w}} N N^{-1} = \g E_1 \g^{-1}\\
\pa_{w} \g_{+} \g_{+}^{-1} &=& 0 \qquad  \qquad
\pa_{w} N_{+} N_{+}^{-1}= 0 \qquad  \qquad
\pa_{w} M M^{-1} = - \g^{-1} E_{-1} \g
\er
as well as 
\be
 \pa_{w} M_{-} M_{-}^{-1} = - \g^{-1}_{-} E_{-1}\g_{-}
\qquad  \qquad
 \pa_{\bar{w}} N_{+} N_{+}^{-1} = -\g^{-1}_{+}  E_1\g_{+}
\lab{nplusfirst}
\ee

So, as a consequence of \rf{pressol} and \rf{gauss}  
the expectation values are
\be
\bra{\l} \g^{-1} \ket{\l} =
\bra{\l} \g_+\({\bar{w}}\) N_{+}\({\bar{w}}\)
M_{-}^{-1}\(w\)\g_{-}^{-1}\(w\) \ket{\l}
\lab{sol}
\ee
This expectation values depends on two parameters 
$\g_{\pm}$. $N_{+}$ and $M_{-}$ are
determined from \rf{nplusfirst}.

Denoting
\be
\g_{+} = \exp\( \theta_{+}\({\bar{w}}\) H^0
+ \xi_{+} \({\bar{w}} \) C\) \qquad \qquad 
\g_{-} = \exp\( \theta_{-}\( w \) H^0
+ \xi_{-} \( w \) C\)
\lab{gammapmdef}
\ee
one gets from \rf{gammadef} and \rf{sol}
\br
e^{ - \nu } &=& e^{ \xi_{+}-\xi_{-}}
\bra{\l_0}  N_{+}\( {\bar{w}}\)  M_{-}^{-1}\(w \) \ket{\l_0} \\
e^{-\vp - \nu }&=&
e^{ \theta_{+} - \theta_{-}  + \xi_{+}-\xi_{-}}
\bra{\l_1}  N_{+}\({\bar{w}} \)  M_{-}^{-1}\( w \) \ket{\l_1}
\er
At this point we are able to write 
the general solution of the model
\br
e^{-\vp} &=&
\frac{\bra{\l_1}  N_{+}\({\bar{w}}\)  M_{-}^{-1}\(w\) \ket{\l_1}}
{\bra{\l_0}  N_{+}\({\bar{w}}\)  M_{-}^{-1}\(w\) \ket{\l_0}} \;
e^{ \theta_{+} - \theta_{-} }
\lab{gensol1}\\
e^{-\nu} &=&
\bra{\l_0}  N_{+}\({\bar{w}} \)  M_{-}^{-1}\(w\) \ket{\l_0} \;
e^{\xi_{+}-\xi_{-}}
\lab{gensol2}
\er

Notice that the
group element $W$ introduced in \rf{wdef} have to  be regular, since
otherwise $A_{\mu}$ will not be flat. Remember that in order for $A_{\mu}$ to
satisfy the zero curvature condition it is necessary that the derivatives
commute when acting on $W$. It then follows that the group elements $g_1$,
$g_2$ and $\g$ have also to be regular.
By regular we mean a quantity such that  derivatives $\pa_{w}$ and 
$\pa_{{\bar  w}}$ 
commute when acting on it. From (see \ct{ggv})
\be
\pa_{\bz} z^{-k-1} = \(-1\)^k \frac{\pi}{k!} \d^{(k,0)}\(\bz,z\)
\lab{powerderiv}
\ee
one observes that
\be
\pa_z \pa_{\bz} z^{-k-1} = \pa_{\bz}\pa_z z^{-k-1}
\ee
Therefore, the fields $\vp$ and
$\nu$, as well as the parameters $\theta_{\pm}$ and $\xi_{\pm}$
can have log singularities, so that the corresponding group elements
will have at
most poles.

The general solution of \rf{phieqmot},
and so of \rf{sinhdelta},
is given by
\rf{gensol1}. Therefore the Riemannian instanton should correspond 
to some particular
choice of the parameters of it. Due to \rf{epm1}, \rf{nplusfirst} 
 and   \rf{gammapmdef} we have
\br
\pa_{\bar w} N_{+} N_{+}^{-1} &=&
- g\,  \( e^{-2 \theta_{+}} T_{+}^0
+ \ba\({\bar w}\) e^{2 \theta_{+}}T_{-}^1\)
\lab{nplus}\\
\pa_{w} M_{-} M_{-}^{-1} &=&
- g\, \( a\( w\) e^{-2 \theta_{-}}T_{+}^{-1}
+  e^{2 \theta_{-}}T_{-}^0 \)
\lab{mminus}
\er
This means that we have a solution to  the model once we 
specify the parameters $\theta_\pm,\, \xi_\pm$, and
solve \rf{nplus}, \rf{mminus} for $M_{-},\, N_{+}$.
\subsection{The Riemannian instanton solutions}

As we have seen, the general solution of the model \rf{phieqmot}, or
equivalently \rf{sinhdelta}, is given by \rf{gensol1}. We now have to
choose the parameters and integration constants of the general
solution, in order to obtain the Riemannian instantons solutions with
the properties described in the introduction. We begin by choosing the
functions $\theta_{\pm}$ as 
\be
\theta_{+} =  -\frac{1}{4} \ln \ba  \qquad \qquad \qquad
\theta_{-} = \frac{1}{4} \ln a
\lab{boundarytheta}
\ee
That choice simplifies the integration of the elements $N_{+}$ and
$M_{-}$. Indeed, \rf{nplus} and \rf{mminus} become
\be
\pa_{\bar w} N_{+} N_{+}^{-1} =
- g\, \sqrt{\ba\(\bar{w} \)} \:\;\:\; b_{1}
\qquad \qquad 
\pa_{w} M_{-} M_{-}^{-1} = - g\,\sqrt{a\( w\)} \:\; \:\;b_{-1}
\lab{b-}
\ee
The operators $b_{1}$ and $b_{-1}$  are elements
of a Heisenberg subalgebra of the $\widehat{sl}(2)$ Kac-Moody
algebra \ct{kac,olive0}. That is like an  algebra
of harmonic oscilators, i.e. they are generated by
\be
b_{2n+1} \equiv  T_{+}^{n} +  T_{-}^{n+1}
\qquad \qquad
\lb b_{2m+1} ,\; b_{2n+1} \rb = C (2m+1) \d _{m+n+1,\,0}
\ee

We can then integrate  \rf{b-}
\br
N_{+} &=& e^{I_{+}\:\; b_{1}}h_{+} \qquad \qquad
I_{+} = -g \int d\bar{w} \: \sqrt{\ba\( \bar{w}\)} 
\; = \; -g \bar{\zeta}
\lab{h+} \nonu\\
M_{-} &=& e^{I_{-} \:\; b_{-1}}h_{-} \qquad \qquad
I_{-} = -g \int dw \: \sqrt{ a\( w\)}  
\; = \;-g \zeta \lab{h-}
\lab{nplusmminusb}
\er
where we used the change of variables to \rf{zeta}, and where $h_{\pm}$ are
constant group elements  obtained by exponetiating the affine
Kac-Moody algebra (integration constants).

We now return  to the general solution \rf{gensol1}. With our choice
of  $\theta_{\pm}$ \rf{boundarytheta} and 
in view of \rf{phiurel} we
get
\be
e^{- u} = \frac{\bra{\l_1}  N_{+}  M_{-}^{-1} \ket{\l_1}}
{\bra{\l_0}  N_{+}  M_{-}^{-1} \ket{\l_0}}
\lab{usol}
\ee

Here we come to a crucial point in the construction of the Riemannian
instanton solution. As it is well known \ct{solspec,fms}, the
one-soliton solutions are 
obtained by taking the integration constants $h_{\pm}$, such that 
$h_{+}h_{-}^{-1} = e^{V(\mu )}$, where $V(\mu )$ is an element of the
Kac-Moody algebra which is an eigenstate of the oscillators
$b_{2n+1}$, i.e.
\be 
\lb  b_{2n+1} ,\; V(\mu) \rb  = - 2 \, \mu^{2n+1}\; V(\mu) 
\lab{eigenv}
\ee
The construction of the operator $V(\mu)$ is explained in appendix
\ref{app:sl2km}. Its expression in terms of a special basis of
the Kac-Moody algebra is given in \rf{vmudef}. However, the nice 
properties of such operator are  best appreciated in the principal vertex
operator representation of the Kac-Moody algebra. Its form in that
representation is given in \rf{normalv}. An important relation
satisfied by $V(\mu)$ is given in \rf{vzvw}. From it we observe that 
\be
V(\mu) V(\nu) \ra 0 \qquad {\rm for} \qquad \mu \ra \nu
\lab{truncation}
\ee
That implies that the exponential $e^{V(\mu )}$ truncates in first
order, and so we do not have convergence problems in our
expressions. Such property is what makes the vertex operator
representation to deserve the name of integral representation
\ct{kac,olive0}. It also explain the truncation of the Hirota's expansion of
the tau functions, since those are nothing more than special expectation
values of $V(\mu)$ in the states of the vertex operator representation
\ct{fms}.  

If one takes $a\(z\)$ to be constants and choose the integration
constants $h_{\pm}$, such that  $h_{+}h_{-}^{-1} = e^{V(\mu )}$, then
one obtains from  \rf{usol} the one-soliton solution of the
sinh-Gordon equation (by taking $u\ra i\, u$ one gets the sine-Gordon
one-soliton). The parameter $\mu$ is related to the rapidity $\theta$
of the soliton through $\mu \equiv \epsilon e^{\theta}$, with
$\epsilon = \pm 1$. It is $\epsilon$ what determines the sign of the
topological charge (in the case of sine-Gordon) and what makes the
difference between the soliton and anti-soliton solutions.  

The $n$-soliton solution is obtained by taking $h_{+}h_{-}^{-1}$ as a
product of those exponentials, i.e. $h_{+}h_{-}^{-1} = \prod_{i=1}^n 
e^{V(\mu_i )}$. As we now explain the  Riemannian instanton solution
is obtained by taking $h_{+}h_{-}^{-1}$ to be a {\rm continuous}
infinite product of exponentials $e^{V(\mu_i )}$. In fact, we shall take the
product in such a way that exponentials for smaller values of $\mu_i$ appear
on the left, and we vary $\mu_i$ continuosly from zero to $+\infty$. In
addition, each value of $\mu_i$ appears only once, without repetition. So,
what we have is an $N$-soliton solution, with $N\ra \infty$, where all the
rapidities appear once, and we do not have a mixture of soliton and
anti-solitons 
since the $\mu_i$'s are all positive. Therefore, we have some sort of soliton
condensate\footnote{We are indebted to Olivier Babelon 
for pointing out to us that
  the Riemannian instanton solution should have such structure. His intuition
  came from his experience with D. Bernard on the calculations of form factors
  and correlation functions for the Ising model \ct{fredholm2}}.

In order to build up the solution we start with an infinite {\em discrete}
product of exponentials, and later take the continuous limit. So, we take the
constant $h_{+},h_{-}$  in 
\rf{h+}, \rf{h-} to be
\be 
h_{+}h_{-}^{-1} \equiv
\prod_{i=1}^{\infty} \; e^{V(\mu_i)} 
\lab{infinitev}
\ee 
From this point on we explore some useful properties of
the vertex operators and its action on the representation
states, which will allow us to end up with a closed expression for
the solution $u$.
So, once  the constants $h_{+}, h_{-}$ have been chosen as in \rf{infinitev},
the solution \rf{usol} depends on
\be 
\bra{\l} N_{+} M_{-}^{-1} \ket{\l} =
\bra{\l} \prod_{i=1}^{\infty} \; (1 + e^{ \b \(\mu_i\)  } \: V(\mu_ i)
)  \ket{\l}  \qquad \qquad 
\b \(\mu_i\) \equiv  2 g \(\mu_i \; {\bar \zeta} +  \, \frac{\zeta}{\mu_i}\) 
\lab{ginto}
\ee
where we have used \rf{nplusmminusb}, \rf{eigenv} and \rf{truncation}. 
Using \rf{nvexpectvalue} and \rf{clpm}, we can expand \rf{ginto} in terms of
sums as  
\br 
\bra{\l_0}  N_{+}  M_{-}^{-1}\ket{\l_0} &=& 
1  \: + \:  \sum_{i}  e^{ \b \(\mu_i\) } \: + \:
\sum_{i< j} e^{ \b \(\mu_i\) } e^{ \b \(\mu_j\) } \; 
\( \frac{ \mu_j - \mu_i}{\mu _j  + \mu_i} \) ^{2}
\nonu\\
&+&  \sum_{i< j <k}  e^{ \b \(\mu_i\) } e^{ \b \(\mu_j\) }  
e^{ \b \(\mu_k\) } \;  \(
\frac{\mu _j - \mu _i}{\mu _j + \mu_i}  \) ^{2} \;  \( \frac{\mu_k 
- \mu _i}{\mu _k + \mu_i}  \) ^{2} \;  \( 
\frac{\mu _k - \mu_j}{\mu _k + \mu_j}  \) ^{2 } + \dots
\nonu\\ & & \nonu\\
\bra{\l_1}  N_{+}  M_{-}^{-1} \ket{\l_1} &=& 1 \: - \: \sum_{i}
e^{ \b \(\mu_i\) } \: + \:  \sum_{i< j}  e^{ \b \(\mu_i\) } 
e^{ \b \(\mu_j\) } \; \(\frac{\mu _j - \mu _i}{\mu _j + \mu_i}  \)
^{2 }  \lab{sumsvev}\\
&-&
 \sum_{i< j <k}  
e^{ \b \(\mu_i\) } e^{ \b \(\mu_j\) }  e^{ \b \(\mu_k\) }
\;  \( \frac{\mu _j - \mu _i}{\mu _j + \mu_i}  \) ^{2} \;  \(
\frac{\mu _k - \mu _i}{\mu _k + \mu_i}  \) ^{2 } \; \( \frac{\mu
_k - \mu _j}{\mu _k + \mu_j}  \) ^{2} \: + \: \dots 
\nonu 
\er
Remember that we take the $\mu_i$'s to be real and positive, and that two 
$\mu_i$'s never coincide.

Expressions \rf{sumsvev} can be written in the form of Fredholm 
determinants. A similar result was found in \ct{fredholm2}, where 
exact correlation functions of the Ising Model were shown to be 
related to the tau functions of the sinh-Gordon model. Following
\ct{fredholm2} we get  
\br
\bra{\l_0}  N_{+}  M_{-}^{-1}\ket{\l_0} &=& {\rm det} \( 1 + W\)\nonu\\
\bra{\l_1}  N_{+}  M_{-}^{-1} \ket{\l_1} &=&{\rm det} \( 1 - W\)
\er
where $W$ is the matrix
\be
W_{ij}\equiv e^{\b\(\mu_i\)/2}  
\frac{\sqrt{4 \mu_i\mu_j}}{\mu_i+\mu_j}e^{\b\(\mu_j\)/2}
\lab{olivierw}
\ee
Using \rf{usol} one then gets
\be
u = \ln\(\frac{{\rm det} \( 1 + W\)}{{\rm det} \( 1 - W\)}\) = 
\Tr \, \ln \frac{1 + W}{1 - W}
\lab{discretesol0} 
\ee
where we used $\ln {\rm det} M = \Tr \ln M$. Expanding the logarithm one gets  
\be
u = 2 \sum_{n=0}^{\infty} \frac{\Tr \; W^{2n+1}}{2n+1}
\lab{discretesol}
\ee

\subsection{The continuous limit}

As we have said, we want to take the limit where the infinite product of
exponentials in \rf{infinitev} becomes a continuous one. In order to do that we
take the label $i$ of the parameter $\mu_i$ to be the rapidity $\theta$ of the
soliton, and let it run from $-\infty$ to $\infty$. We then have that $\mu_i
\ra \mu_{\theta}=e^{\theta}$. Then 
\be
\sum_{i} \ra \L \int_{-\infty}^{\infty} d\theta = \L \int_0^{\infty}
\frac{d\mu}{\mu}  
\lab{continuoussums}
\ee
where $\L$ is a scaling factor of the integration measure. 

Then, from \rf{olivierw} it follows that\footnote{Notice that 
  the subindices of $\mu$  have a different meaning now. They label the
  variables giving the values of rows and columns of the matrices, and not the
  actual values of those as before.} 
\be
\Tr W^N \ra \L^N \int_0^{\infty}\frac{d\mu_{1}}{\mu_{1}} \ldots 
\int_0^{\infty}\frac{d\mu_{N}}{\mu_{N}} 
e^{\b\(\mu_1\)/2}  \frac{\sqrt{4 \mu_1\mu_2}}{\mu_1+\mu_2}e^{\b\(\mu_2\)}\, 
 \frac{\sqrt{4 \mu_2\mu_3}}{\mu_2+\mu_3}e^{\b\(\mu_3\)} \ldots 
e^{\b\(\mu_N\)}  \frac{\sqrt{4 \mu_N\mu_1}}{\mu_N+\mu_1}e^{\b\(\mu_1\)/2}
\ee
Therefore \rf{discretesol} becomes
\be
u = 2 \sum_{n=0}^{\infty} \frac{\(2\L\)^{2n+1}}{2n+1}\, I_{2n+1}
\lab{continuoussol}
\ee
with
\be
I_{N}\equiv \int_0^{\infty}\frac{d\mu_{1}}{\mu_{1}} \ldots 
\int_0^{\infty}\frac{d\mu_{N}}{\mu_{N}}
\frac{\mu_1}{\(\mu_1+\mu_2\)}\frac{\mu_2}{\(\mu_2+\mu_3\)} \ldots 
\frac{\mu_N}{\(\mu_N+\mu_1\)} \, e^{\b\(\mu_1\) + \ldots +\b\(\mu_N\)}
\lab{indef}
\ee
In the case $N=1$ we have that
\br
I_1 &=& \h\, \int_{0}^{\infty}\, \frac{d\mu}{\mu} e^{\b\(\mu\)} = 
 K_0 \( 4 \mid g\mid \mid \zeta\mid\)
\lab{i1def}
\er
where $K_0$ is the modified Bessel function. In fact the above
  expression is valid for ${\rm Re} \(g\zeta\) <0$. However, we shall
  take it to be valid for ${\rm Re} \(g\zeta\) >0$ too (see comments below
  \rf{fpnfmnexp}).

Notice that the integrals $I_N$ are real. Indeed, from \rf{ginto} we have 
\be
\b^* \(\mu_i\) = \b \(\frac{1}{\mu_i}\)
\ee
Therefore, one can undo the complex conjugation with the change of
integration variables,  $\mu_i \ra 1/\mu_i$, since 
$\int_0^{\infty}\frac{d\mu_{i}}{\mu_{i}}$ is left unchanged. In
addition,  
 $\mu_i/\(\mu_i+\mu_j\) \ra \mu_j/\(\mu_i+\mu_j\)$, and so
the product of those terms in the integrand of \rf{indef} is left
invariant. Consequently,  the solution $u$ given
in \rf{continuoussol} is real (since $\L$ is real). 

We now want to analyse the boundary conditions satisfied by the
solution \rf{continuoussol}. In order to do that we perform the change
of integration variables
\br
\phi_i &\equiv& \ln \frac{\mu_i}{\mu_{i+1}} \qquad \qquad i=1,2,\ldots
N-1
\nonu\\
\nu &\equiv & \( \prod_{i=1}^N \mu_i\)^{1/N}
\er
The integrals \rf{indef} become 
\br
I_{N} &=& \frac{1}{2^N}\; \int_{-\infty}^{\infty}\, d\phi_1 \ldots 
\, d\phi_{N-1} 
\,\frac{1}{\cosh \(\h\phi_1\)\cosh\(\h\phi_2\)\ldots\cosh\(\h\phi_{N-1}\)
\cosh\(\h\sum_{n=1}^{N-1}\phi_{n}\)} \times\nonu\\
&\times& 
\int_0^{\infty} \frac{d\nu}{\nu}\; 
e^{2 g \( {\bar \zeta} f_N\(\phi\)\; \nu 
+ \zeta  f_N\(-\phi\)\; \frac{1}{\nu}\)}
\lab{irregint}
\er
where
\be
f_N\(\phi\) \equiv  \sum_{l=1}^N \exp\( 
\frac{1}{N}\( - \sum_{n=1}^{l-1} n \, \phi_n + 
\sum_{n=l}^{N-1} \(N-n\) \, \phi_n\)\)
\lab{fndef}
\ee
If ${\rm Re} \(g\zeta\) <0$, the integral in $\nu$ in \rf{irregint}
is the modified Bessel function
$K_0$, and so one gets  
\be
I_N = 
\frac{1}{2^{N-1}}\; \int_{-\infty}^{\infty}\, 
d\phi_1 \ldots \, d\phi_{N-1}
\frac{K_0\( 4 \mid g\mid \mid\zeta\mid \sqrt{w_N}\)}
{\cosh \(\h\phi_1\)\cosh\(\h\phi_2\)\ldots\cosh\(\h\phi_{N-1}\)
\cosh\(\h\sum_{n=1}^{N-1}\phi_{n}\)}
\lab{finalin}
\ee
where $w_N$ is given by
\be
w_N \equiv f_N\(\phi\) f_N\(-\phi\) = N + 
2  \sum_{l=0}^{N-2} \sum_{j=1}^{N-l-1}\cosh \sum_{i=j}^{j+l} \phi_i
\lab{fpnfmnexp}
\ee
However, we shall take the expression \rf{finalin} to be also valid for
${\rm Re} \(g\zeta\) >0$. Such analytical continuation process will
be justified later when we shall check the validity of the solution by
directly replacing it into the equations of motion. Therefore, the
solution \rf{continuoussol} depends  on $\zeta$ and $g$  through
their norms only.  

\subsection{The boundary conditions}

For large arguments the modified Bessel function $K_0$ have the 
following behaviour 
\be
K_0\(z\) \sim \sqrt{\frac{\pi}{2z}} \, e^{-z} \( 1 + O\(\frac{1}{z}\) \)
\qquad \qquad\qquad \mbox{\rm for large $z$}
\lab{largez}
\ee
Consequently it is clear that
\be
I_N \ra 0 \qquad \qquad {\rm for} \qquad \mid\zeta\mid \ra \infty
\lab{inatinfinty}
\ee
and so, the $u$ field does go to zero for large $\zeta$, as required
(see \rf{boundarycond2}). 

The analysis for small $\zeta$ is trickier. The reason is that taking
$\mid\zeta\mid$ small does not guarantee that the argument of the Bessel
function $K_0$ is small, since $w_N$
can be infinitely
large. However, in the region where $\mid\zeta\mid\, w_N$
diverges for small $\mid\zeta\mid$ the function $K_0$ vanishes and so there is
no contribution for the integral $I_N$. Therefore, 
we can use  the following reasoning: Let $\mid\zeta\mid$ have a
fixed infinitesimal value $\mid\zeta\mid = \varepsilon$. We split the domain of
integration in two regions, namely
\br
D_0 \equiv  \mbox{\rm region of $\( \phi_1, \ldots , \phi_{N-1}\)$ where 
$\varepsilon 4 \mid g\mid \sqrt{w_N} <
\sqrt{\varepsilon}$} \nonu\\  
D_1 \equiv  \mbox{\rm region of $\( \phi_1, \ldots , \phi_{N-1}\)$ where 
$\varepsilon 4 \mid g\mid \sqrt{w_N} > \sqrt{\varepsilon}$}
\er
In the region $D_0$ we use the fact that for small arguments,  $K_0$
diverges as
\be
K_0\(z\) \sim - \ln \frac{z}{2}\( 1 + O\(z^2\)\)   \qquad \qquad
\mbox{\rm for small $z$}
\ee
and so
\br
I_{N} &=& \frac{1}{2^{N-1}}\; \int_{D_0}\, d\phi_1 \ldots 
\, d\phi_{N-1} 
\,\frac{-\ln \( 2 \mid g\mid \mid\zeta\mid \sqrt{w_N}\) }
{\cosh \(\h\phi_1\)\cosh\(\h\phi_2\)\ldots\cosh\(\h\phi_{N-1}\)
\cosh\(\h\sum_{n=1}^{N-1}\phi_{n}\)} \nonu\\
&+&\frac{1}{2^{N-1}}\; \int_{D_1}\, d\phi_1 \ldots 
\, d\phi_{N-1} 
\,\frac{K_0\( 4 \mid g\mid \mid\zeta\mid \sqrt{w_N}\)} 
{\cosh \(\h\phi_1\)\cosh\(\h\phi_2\)\ldots\cosh\(\h\phi_{N-1}\)
\cosh\(\h\sum_{n=1}^{N-1}\phi_{n}\)}\nonu\\
\lab{splitint}
\er

Notice that in the region $D_1$ the argument of $K_0$ never vanishes
and so $K_0$ is finite there. On the other hand, on $D_1$ we
have  
\be
w_N > \frac{1}{4 \mid g\mid\sqrt{\varepsilon}}
\ee
and so $w_N \ra \infty$ as $\varepsilon \ra 0$. But from
\rf{fpnfmnexp} one observes that the only way for that to happens is
that at least one the $\phi_i$'s should diverge. Therefore, the denominator
of the integrand of \rf{splitint}, in the $D_1$ region, diverges. So,
one gets that the integral in $D_1$ in \rf{splitint} vanishes for
$\varepsilon \ra 0$.

Consequently, the integral in $D_0$ in \rf{splitint} implies that  
\be
I_{N} \sim -\; \kappa_N\;\ln\mid\zeta\mid 
\qquad \qquad {\rm for} \qquad \mid\zeta\mid \ra 0
\lab{hiperten}
\ee
with
\br
\kappa_N \equiv 
\frac{1}{2^{N-1}}\; \int_{-\infty}^{\infty}\, 
\,\frac{d\phi_1 \ldots \, d\phi_{N-1} }
{\cosh \(\h\phi_1\)\cosh\(\h\phi_2\)\ldots\cosh\(\h\phi_{N-1}\)
\cosh\(\h\sum_{n=1}^{N-1}\phi_{n}\)}
\nonu\\
\lab{kappandef}
\er 
Performing the integration using the fact that 
\br
\kappa_N&=&
\int_{-\infty}^{\infty}\, d\phi_1 \ldots 
\, d\phi_{N} \;\frac{\d\( \sum_{i=1}^{N}
  \phi_i\)}{\prod_{i=1}^{N}\cosh \phi_i}
=\frac{1}{2\pi}\, \int_{-\infty}^{\infty}\, dk \, 
\int_{-\infty}^{\infty}\, d\phi_1 \ldots 
\, d\phi_{N} \;\frac{e^{ik \sum_{i=1}^{N} \phi_i}}{\prod_{i=1}^{N}\cosh
\phi_i}\nonu\\
&=&
\frac{1}{2\pi}\, \int_{-\infty}^{\infty}\, dk \,
\( \int_{-\infty}^{\infty}\, d\phi \frac{e^{ik \phi}}{\cosh\phi}\)^N
\er
one gets
\be
\kappa_{2n+1} = \frac{\pi^{2n}}{2^n} \frac{\(2n-1\)!!}{n!}
\lab{kappnsolved}
\ee
Consequently we have from \rf{continuoussol} that 
\be
u \sim -\frac{2}{\pi}\, f\( 2\pi\L\) \, \ln \mid \zeta \mid 
\qquad \qquad {\rm for} \qquad \mid\zeta\mid \ra 0
\lab{uatzero}
\ee
where
\be
f\(x\)\equiv \sum_{n=0}^{\infty} \frac{\(2n-1\)!!}{2^n n!\(2n+1\)}
\; x^{2n+1}
\lab{fdef}
\ee
Notice this series is convergent for $x^2<1$, 
and divergent for $x^2>1$. For $x^2=1$ the ratio 
test does not say anything, but  
one can check that it does converge there and $f\(1\)=
\pi/2$. Therefore, we must have $\mid \L\mid \leq 1/2\pi$.

Our function $a(z)$ is supposed to be a polynomial and to represent
a hyperelliptic Riemann surface.
As discussed in Appendix \ref{app:delta}, 
a good local coordinate near a branch 
point $z_i$ is $\xi_i= \sqrt{z-z_i}$, i.e., near $z_i$
we have $z= z_i + {\xi_i}^2$. So, according to \rf{comportamento}, 
near the branching cut, we must have $\zeta \sim a^{3/2}$.
Therefore, 
\be
u \sim -\frac{3}{\pi}\, f\( 2\pi\L\) \,\ln\mid a\mid
\ee

One can check that 
\be
f\( \h\)= \frac{\pi}{6}
\lab{nicef}
\ee
Consequently, in order to satisfy the boundary condition
\rf{boundarycond}, we must set  
\be
\L = \frac{1}{4\pi} 
\lab{niceL}
\ee

Therefore from \rf{continuoussol}, the desired solution to
\rf{sinhdelta} is given by 
\be
u = 2 \sum_{n=0}^{\infty} \frac{I_{2n+1}}{\(2n+1\)\(2\pi\)^{2n+1}}
\lab{finalsol}
\ee
where $I_{2n+1}$ is given in \rf{i1def} and \rf{finalin}. 
Consequently, it depends only on the combination 
$4 \mid g\mid \mid\zeta\mid$,  and it is symmetric, on the plane $\zeta$, under
rotation around the origin $\zeta =0$.

\section{Check of the solution}
\label{sec:check}

As we have seen in \rf{niceL}, the value of the scaling factor $\L$ of
the integration measure, introduced in \rf{continuoussums}, was fixed
to $1/4\pi$. That was imposed by the behaviour of the solution at
$\zeta =0$ (or equivalently on the zeroes of $a\(z\)$). However,
as we will see in this section, the solution holds true for any value
of $\L$, outside the zeroes of $a\(z\)$. That is a consequence of very
interesting non-linear differential equations satisfied by the
integrals $I_{2n+1}$ defined in \rf{finalin}. Therefore, in order to
emphasize those properties, we shall not fix $\L$ in this check
of the solution. 

We begin by looking at the convergence of the series
\rf{continuoussol}. As we have seen in \rf{inatinfinty}, the integrals
$I_{2n+1}$ go to zero for large arguments. In fact, from the numerical
data in tables  \ref{tab:i1} and \ref{tab:i3i5i7} of appendix
\ref{sec:nume}, one sees that they 
decay faster for larger index $2n+1$. Therefore, we should not have
problems of convergence of the series \rf{continuoussol} for large
arguments. For $\zeta$ close to zero one can use \rf{hiperten} and
\rf{kappnsolved} and the ratio test to check that the series converges
for $\L < \sqrt{2}/2\pi$. Therefore, the series \rf{finalsol} for the
final solution should converge everywhere. 

We begin by calculating the derivatives of the field $u$ close to
$\zeta =0$. From \rf{uatzero} it follows that
\be
\pa_{\zeta} \, u \sim -\frac{1}{\pi}\, f\( 2\pi\L\) \, \frac{1}{\zeta}
\ee
and so using \rf{powerderiv} 
\be
\pa_{{\bar\zeta}}\pa_{\zeta} u \sim - f\( 2\pi\L\) \, 
\d\(\zeta ,{\bar\zeta}\) \,
\lab{deriu}
\ee
Some care must be taken here since we have been working with delta 
functions in different coordinate frames. In order avoid 
misunderstandings which can lead to inconsistencies in fixing 
$\L$, we devote a discussion on this point in Appendix 
\ref{app:delta}. From \rf{mudadelta} we see that
\be
\pa_{{\bar\zeta}}\pa_{\zeta} u 
\sim - f\( 2\pi\L\) \, \d\(\zeta ,{\bar\zeta}\) = 
- f\( 2\pi\L\) \,\frac32 \, \d \( a , {\bar{a}}\) \:
\pa_\zeta a \, \pa_{\bar{\zeta}} {\bar{a}} \,,
\lab{rightbc1}
\ee
and using \rf{nicef} and \rf{niceL} we get
\be
\pa_{{\bar\zeta}}\pa_{\zeta} u 
\sim - \frac{\pi}{6} \, \d\(\zeta ,{\bar\zeta}\) = 
- \frac{\pi}{4} \, \d \( a , {\bar{a}}\) \:
\pa_\zeta a \, \pa_{\bar{\zeta}} {\bar{a}} \,.
\lab{rightbc0}
\ee
Therefore, we do have that \rf{sinhdelta} is satisfied at $\zeta =0$. 

Let us now evaluate the derivatives of $u$ for $\zeta \neq 0$. Since
$I_{2n+1}$ depends on $\zeta$ through the modified Bessel function
$K_0$, we consider 
\br
\pa_{{\bar{\zeta}}} \pa_{\zeta} 
K_0 \( 4 \mid g\mid \mid\zeta\mid \sqrt{w_N}\)&=&
\(4 \mid g\mid\)^2 w_N \; 
\pa_{{\bar{\zeta}}} \mid \zeta\mid \;\pa_{\zeta} \mid \zeta\mid \; 
K_0^{\pr\pr}\( 4 \mid g\mid \mid\zeta\mid \sqrt{w_N}\)
\nonu\\
&+&
\,4 \mid g\mid \sqrt{w_N} \; \pa_{{\bar{\zeta}}}\pa_{\zeta} \mid \zeta\mid \; 
K_0^{\pr}\( 4 \mid g\mid \mid\zeta\mid \sqrt{w_N}\)
\er
Observe that 
\br
\pa_{\zeta} \mid \zeta\mid = 
\h \sqrt{\frac{{\bar{\zeta}}}{\zeta}} \;\;\;\;, \qquad 
\pa_{{\bar{\zeta}}}\mid \zeta\mid &=& \h \sqrt{\frac{\zeta}{{\bar{\zeta}}}}
\lab{remark1}
\er
and with help of \rf{powerderiv} we obtain
\be
\pa_{{\bar{\zeta}}}\pa_{\zeta} \mid \zeta\mid = 
\frac{1}{4 \mid \zeta\mid} + 
\frac{\pi}{4}  \mid \zeta\mid\d \( \zeta , {\bar{\zeta}}\)
\lab{remark2}
\ee
Since we are taking the point $\zeta =0$ out, we can
use the defining equation of $K_0$ namely, 
\be
z^2 K_0^{\pr\pr}\(z\) +z K_0^{\pr}\(z\) -z^2 K_0\(z\) =0
\lab{besseleq}
\ee
and  \rf{remark1} and \rf{remark2} to get 
\br
\pa_{{\bar{\zeta}}} \pa_{\zeta} 
K_0 \( 4 \mid g\mid \mid\zeta\mid \sqrt{w_N}\)=
4 g^2 w_N \; K_0\( 4 \mid g\mid \mid\zeta\mid \sqrt{w_N}\)
\lab{preinjnrel}
\er

Therefore, from \rf{finalin} one has
\be
\pa_{{\bar{\zeta}}} \pa_{\zeta}I_{N} = 4 g^2 \, J_N \qquad \qquad \zeta \neq 0
\lab{injnrel}
\ee
where 
\be
J_N \equiv 
\frac{1}{2^{N-1}}\; \int_{-\infty}^{\infty}\, 
d\phi_1 \ldots \, d\phi_{N-1}
\frac{w_N \; K_0\( 4 \mid g\mid \mid\zeta\mid \sqrt{w_N}\)}
{\cosh \(\h\phi_1\)\cosh\(\h\phi_2\)\ldots\cosh\(\h\phi_{N-1}\)
\cosh\(\h\sum_{n=1}^{N-1}\phi_{n}\)}
\lab{jndef}
\ee
Notice that the relation \rf{injnrel} is also valid for $N=1$, with 
\be
J_1 = K_0 \( 4 \mid g\mid \mid \zeta\mid\)
\lab{j1def}
\ee
The reason is that from
\rf{i1def} we have that $I_1 = K_0 \( 4 \mid g\mid \mid \zeta\mid\)$,
and so  \rf{preinjnrel} becomes \rf{injnrel}  using the fact that $w_1=1$. 

Using arguments similar to those leading to \rf{preinjnrel}, one can check
that $I_N$ and $J_N$ satisfy
\be
z^2 I_N^{\pr\pr} \(z\) + z I_N^{\pr} \(z\) = z^2 J_N\(z\)
\lab{niceinjnrel}
\ee
where $z$ stands for the argument of those functions, i.e. $z\equiv 4 \mid
g\mid \mid \zeta\mid$.

We now have, from \rf{continuoussol}, \rf{injnrel} and \rf{rightbc1}, that
\be
\pa_{{\bar{\zeta}}} \pa_{\zeta}\, u = 8 g^2 
\sum_{n=0}^{\infty} \frac{\( 2 \L\)^{2n+1}}{2n+1} \, J_{2n+1} - 
\d \( \zeta , {\bar{\zeta}}\)\; f\(2\pi \L\)
\ee
where $f\(x\)$ was defined in \rf{fdef}.

So, replacing into equation \rf{sinhdelta},  we get
\be
2 g^2\, \( 
4\, \sum_{n=0}^{\infty} \frac{\( 2 \L\)^{2n+1}}{2n+1} \, J_{2n+1} - 
\sinh 2 u \) =  
 \d \( \zeta , {\bar{\zeta}}\)\; f\(2\pi \L\)
-\frac{\pi}{4} \d\( a\) \pa_{\zeta} a \pa_{{\bar \zeta}} \ba 
\lab{zero=zero}
\ee
As we have already seen the vanishing of the r.h.s. of \rf{zero=zero}
fixes the value of $\L$. Indeed,  using \rf{mudadelta}  we see we 
need to choose $\L$
such that $f\(2\pi \L\)=\pi/6$. But that is exactly what we had done in
\rf{nicef} and \rf{niceL} to get the right boundary conditions.  

The l.h.s. of \rf{zero=zero} on the other hand vanishes for any $\L$. That is
an amazing result and involves special properties of the Bessel
function $K_0$, or more precisely of $I_N$ and $J_N$, which we could
not find in the literature.   Expanding the
l.h.s. of \rf{zero=zero} in powers of $\L$ we get
\br
\L & \ra & K_0 \( 4 \mid g\mid \mid \zeta\mid\) = 
K_0 \( 4 \mid g\mid \mid \zeta\mid\) \qquad \( I_1=J_1\) \nonu\\
\L^3 & \ra & J_3 = I_3 + 8 I_1^3 
\lab{curiousrel}\\
\L^5 & \ra & J_5 = I_5 + \frac{40}{3} I_1^2 \, I_3 + \frac{32}{3} I_1^5
\nonu\\  
\L^7 & \ra & J_7 = I_7 + \frac{224}{9} I_1^4 \, I_3 + 
\frac{56}{9} I_1 \, I_3^2 + \frac{56}{5} I_1^2 \, I_5 + 
\frac{256}{45} I_1^7 \nonu\\
\vdots & \vdots & \vdots \nonu 
\er
With help of \rf{niceinjnrel} we get that the integrals $I_{2n+1}$ 
satisfy the following coupled non-linear differential equations  
\br
I_3^{\pr\pr} + \frac{1}{x} I_3^{\pr} - I_3 &=&  8 I_1^3 
\nonu\\
I_5^{\pr\pr} + \frac{1}{x} I_5^{\pr} - 
I_5 &=& \frac{40}{3} I_1^2 \, I_3 + \frac{32}{3} I_1^5 \nonu\\
I_7^{\pr\pr} + \frac{1}{x} I_7^{\pr} -
I_7 &=& \frac{224}{9} I_1^4 \, I_3 + 
\frac{56}{9} I_1 \, I_3^2 + \frac{56}{5} I_1^2 \, I_5 + 
\frac{256}{45} I_1^7 
\lab{nldeq}\\
\vdots & \vdots & \vdots \nonu
\er
So, the r.h.s. of these equations is what makes the difference between 
$K_0\(x\)$ and the $I_{2n+1}$'s (see \rf{besseleq}). 

We did not find a way of checking relations \rf{nldeq}
analytically. In the appendix \ref{sec:nume} we make a  
  numerical check of them up to the equation for $I_7$, and find that
  they are indeed true.  

Therefore, the configuration \rf{finalsol} is indeed a solution of
\rf{sinhdelta}.

\vspace{1 cm}

\noindent {\large{\bf Acknowledgements}}\\

We are very grateful to Olivier Babelon for many elucidating discussions
which were crucial in the completion of this work. LAF is partially
supported by CNPq (Brazil), EEL is supported by a scholarship from
FAPESP (Brazil). CPC is grateful for the hospitality at the LPTHE in Paris 
where part of this work was developed,  
and for the agreement CAPES/COFECUB for a grant.

\appendix

\section{Hyperelliptic Riemann surfaces and delta function}
\label{app:delta}

Let us consider the  equation for an hyperelliptic Riemann surface  
\be
y^2 = a(z) = (z-z_1)(z-z_2)...(z-z_n)
\lab{eq.(1)}
\ee
There are branch points at $z=z_1,...,z=z_n$. Let us suppose they are
all distinct. Also the point at infinity is a branch point when 
$n$ is odd. For simplicity let us suppose that $n$ is even.
There are two sheets, which are two
copies of the $z$-plane. Now 
we draw a cut from $z_1$ to $z_2$, another from $z_3$ to $z_4$ and 
so on. The two sheets are now attached to each other through the 
cuts. By proceeding along a small circle around a branch point
we will pass from one sheet to another and after $4\pi$ we are 
back to the 
initial point. Passing to the string interpretation, it is evident 
that now we have a string of length $4\pi$ instead of $2\pi$
like initially. The string interpretation is as follows: we have
initially two strings that interact successively $n-2$ times
and finally split into two separate strings as in the initial state.
The Riemann surface we get in this way is an hyperelliptic
one with two punctures representing the initial strings, two 
representing the final strings and $(n-2)/2$ handles. Let us 
call it $\Sigma$.

Let us return to \rf{eq.(1)}. $y$ and $z$ are coordinates of two complex 
planes, but, of course they can be considered as function over
$\Sigma$. The coordinate $z$ is not a good coordinate near a
branch point. A good local coordinate near a branch 
point $z_i$ is $\xi_i= \sqrt{z-z_i}$. I.e., near $z_i$
we have $z= z_i + {\xi_i}^2$. $z$ is not a good coordinate at 
infinity either, it must be replaced by $w= 1/z$. After these 
substitution we see that $y$ is a meromorphic function, with 
$n$ zeroes at the branch points and a pole of order $n$ 
at $z=\infty$ on each sheet. 

Let us now consider the differential $dz$. $dz \sim \xi_i d\xi_i$
near $z_i$, therefore $dz$ has simple zeroes at the branch points.
At infinity $dz\sim  w^{-2} dw$, therefore it has a double pole
there, on both sheets. 

Therefore the product $ydz$ is a meromorphic one--form over
$\Sigma$, with a single pole of order $n+2$ at infinity.
It makes sense to integrate this form along a path,
and this is what we do when we write $\zeta= \int \sqrt a dz$,
eq.(2.47).

Let us now come to eq. \rf{uatzero} 
\be
u \sim - {2\over \pi} f \, {\rm ln} |\zeta|
\lab{hjk}
\ee
and to the behaviour
\be
u = - {1 \over 2}{\rm ln} |a| 
\lab{ijk}
\ee
which is required for consistency near a branch point. 
Near a generic branch point $a\sim z-z_0$, a good coordinate
is $\xi = \sqrt a$. In terms of this coordinate
we have
\be
a \sim \xi^2 ,\quad\quad \zeta = \int dz \sqrt a \sim \int \xi^2 d\xi
\sim \xi^3, \quad\quad a\sim \zeta^{2/3}
\lab{comportamento}
\ee

Therefore, in order that \rf{hjk}  be consistent with \rf{ijk} we must have 
$f= {\pi \over 6}$.

Some attention must be payed to the definition of the delta
functions, in order to avoid possible inconsistencies
in fixing the value of $f$. Let us see this point in detail.

Consider the good coordinate $\xi$ and
\be
\int d^2 \xi\, \partial_\xi \partial_{\bar\xi} {\rm \ln} |\xi|=
{\pi\over 2} \int d^2 \xi \, \delta(\xi,\bar\xi)
\ee
Since the contribution to the integral is only at the origin we can
restrict it to the unit disk around the origin, and proceed in 
another way by applying Stokes theorem
\be
\int d^2 \xi\, \partial_\xi \partial_{\bar\xi} {\rm \ln} |\xi|=
{1\over 2} \int d^2 \xi\,\partial_{\bar \xi} {1\over \xi} =
{1\over 2} \oint d\xi {1\over \xi} = {1\over 2} \int_0^{2\pi} d\theta= \pi i
\ee
where the contour integral extends over the unit circle 
around the origin and $\xi= e^{i\theta}$.

If we repeat the same calculation with the 'bad' coordinate $a$
we get
\be
\int d^2 a\, \partial_a \partial_{\bar a} {\rm \ln} |a|=
{\pi \over 2} \int d^2 a \, \delta(a,\bar a)
\ee
and 
\be
\int d^2 a\, \partial_a \partial_{\bar a} {\rm \ln} |a|=
{1\over 2} \int d^2 a\,\partial_{\bar a} {1\over a} =
{1\over 2} \oint da {1\over a} = {{4\pi i}\over 2} = 2 \pi i
\ee
The last steps are due to the fact that the angular integration
for $a$ extends over $4\pi$ since $a\sim \xi^2$.

In a similar way for $\zeta$ we will get
\be
\int d^2 \zeta\, \partial_\zeta \partial_{\bar\zeta} {\rm \ln} |\zeta|=
{1\over 2} \int d^2 \zeta\,\partial_{\bar \zeta} {1\over \zeta} =
{1\over 2} \oint d\zeta {1\over \zeta} = {1\over 2} \int_0^{6\pi} d\theta=
3 \pi i
\ee

At this point it is judicious to make use of different symbols for 
these delta functions: $\delta (\xi,\bar\xi)$, which is the usual delta function,
and $\delta_a(a,\bar a), \delta_\zeta(\zeta,\bar \zeta)$ so that,
roughly speaking,
\be
\delta_a(a,\bar a)\sim 2 \delta (\xi,\bar\xi),
\quad\quad \delta_\zeta(\zeta,\bar \zeta)\sim 3 \delta (\xi,\bar\xi)
\ee
In additon we must take into account the Jacobian factor due to 
the change of coordinates (a delta function transforms like the component of a 
one-form). In conclusion we have the relation
\be
\delta_\zeta(\zeta,\bar \zeta)= {3\over 2} \delta_a(a,\bar a)
\partial_\zeta a \partial_{\bar\zeta} \bar a
\lab{mudadelta}  
\ee

\section{The affine $\widehat{sl}(2)$ Kac-Moody algebra}
\label{app:sl2km}

The commutation relations of the affine $\widehat{sl}(2)$ Kac-Moody
algebra are given by
\br
\sbr{H^m}{H^n}&=& 2 C m \d_{m+n,0} \nonu\\
\sbr{H^m}{T_{\pm}^n}&=& \pm T_{\pm}^{m+n}\nonu\\
\sbr{T_{+}^m}{T_{-}^n}&=& H^{m+n} + C m \d_{m+n,0} \nonu\\
\sbr{D}{H^m} &=& m H^m\nonu\\
\sbr{D}{T_{\pm}^m}&=& m T_{\pm}^{m}
\lab{sl2kmcomrel}
\er
In a highest weight representation we can take
\be
\( H^n\)^{\dagger} = H^{-n} \qquad \qquad
\( T_{+}^n\)^{\dagger} = T_{-}^{-n}
\ee
Let $\ket{\l}$ be a highest weight state. Then
\be
T_{+}^m \ket{\l} = T_{-}^n \ket{\l} = H^n\ket{\l} = 0 \qquad \qquad
m\geq 0 \; ; \quad n >0
\ee
Therefore, the norm of the state $T_{+}^{-m} \ket{\l}$, $m>0$,  is
\br
\bra{\l} T_{-}^{m}T_{+}^{-m} \ket{\l}&=&
\bra{\l} \sbr{T_{-}^{m}}{T_{+}^{-m}} \ket{\l} =
\bra{\l}  - H^0 + m C \ket{\l}
\er
Using \rf{l0action} and \rf{l1action}, one gets
\be
T_{+}^{-1} \ket{\l_1} = 0
\ee
Analogously, the norm of the state $T_{+}^{-m}T_{+}^{-m} \ket{\l}$, $m>0$, is
\br
\bra{\l} T_{-}^{m}T_{-}^{m}T_{+}^{-m}T_{+}^{-m} \ket{\l}=
2 \bra{\l}\( - H^0 + m C\) \( - H^0 + m C\) - \( - H^0 + m C\)\ket{\l}\nonu
\er
Therefore, from \rf{l0action}
\be
T_{+}^{-1}T_{+}^{-1} \ket{\l_0}=0
\ee
Using similar arguments one gets that
\be
T_{-}^0 \ket{\l_0} = 0 \qquad \qquad
T_{-}^0 T_{-}^0 \ket{\l_1} = 0
\ee
For the applications we make in this paper it is useful to work with 
following basis for the affine $sl(2)$ Kac-Moody algebra
\br
b_{2m+1} &=& T_{+}^m + T_{-}^{m+1} \lab{b2n}\\
F_{2m+1} &=& T_{+}^m -T_{-}^{m+1} \lab{fn1} \\
F_{2m} &=& H^{m} - {1\o 2} C \d_{m,0} \lab{fn2}
\er
in addition to the generators $C$ (central term) and $D$ .
The commutation relations are
\br
\lb b_{2m+1} \, , \, b_{2n+1} \rb &=& C (2m+1) \d_{m+n+1,0} \lab{hei1} \\
\lb b_{2m+1} \, , \, F_{2n+1} \rb &=& - 2 F_{2(m+n+1)} \lab{comu1} \\
\lb b_{2m+1} \, , \, F_{2n} \rb &=& - 2 F_{2(m+n)+1} \lab{comu2} \\
\lb F_{2m+1} \, , \, F_{2n+1} \rb &=& - C (2m+1) \d_{m+n+1,0} \lab{comu3} \\
\lb F_{2m+1} \, , \, F_{2n} \rb &=& - 2 b_{2(m+n)+1} \lab{comu4} \\
\lb F_{2m} \, , \, F_{2n} \rb &=& C 2 m \d_{m+n,0} \lab{hei2}
\er
The elements which diagonalize the adjoint action of the oscillators
$b_{2m+1}$ are\footnote{We choose the normalization factor to be $-2$ because
  we want the expectation values in \rf{expectvaluev} to be $\pm1$.}
\be
V\( \mu\) = -2 \sum_{n=-\infty}^{\infty} \mu^{-n} F_n
\lab{vmudef}
\ee
Indeed
\be
\sbr{b_{2m+1}}{V\( \mu\)}= -2 \mu^{2m+1}\; V\( \mu\)
\ee

Notice that the expectation value of such operator on a highest weight state
gets contribution only from its zero mode. Indeed, using  \rf{l0action} and
\rf{l1action}, one gets 
\be
\bra{\l_0} V\( \mu\) \ket{\l_0} = 1 \qquad \qquad 
\bra{\l_1} V\( \mu\) \ket{\l_1} = -1
\lab{expectvaluev}
\ee

\subsection{Vertex Operator Construction}
\label{app:vertex}

We now discuss the principal vertex operator representation of the
$\widehat{sl}(2)$ 
Kac-Moody algebra. It differs from the homogeneous vertex representation,
i.e. the Fubini-Veneziano vertex, in the sense that the oscillators have only
odd indices and so do not have zero modes. We introduce 
\br 
Q(z) &=& \sum_{n = 0}^{\infty} \left( \frac{z^{-
\mathbf{N}}}{\mathbf{N}} b_{\mathbf{N}} - \frac{z^{
\mathbf{N}}}{\mathbf{N}} b_{- \mathbf{N}} \right) \qquad \qquad
{\mathbf{N}} = 2n+1 \lab{qpm}\\
P(z) &=& - \sum_{n = 0}^{\infty} \left( {z^{- \mathbf{N}}}
b_{\mathbf{N}} + {z^{ \mathbf{N}}} b_{- \mathbf{N}} \right)
\er
where the oscillators satisfy the commutation relations \rf{hei1}. 
We then have
\br P(z)Q(w) =  \: :: P(z)Q(w) :: \: + \: \frac{wz}{z^2 - w^2} \;
C \qquad \mid z \mid > \mid w \mid \nonu\\
Q(w)P(z) =  \: :: Q(w)P(z) :: \: - \: \frac{wz}{w^2 - z^2} \; C
\qquad \mid w \mid > \mid z \mid \er
We now introduce the vertex operator 
\be \cv (z) = \: :: e^{\a Q(z)}::
\lab{vertexdef}
\ee
and so 
\br P(z)\cv(w) =  \: :: P(z)\cv(w) :: \: + \: \frac{\a \, wz}{z^2 -
w^2} \; C \, \cv(w)
\qquad \mid z \mid > \mid w \mid \nonu\\
\cv(w)P(z) =  \: :: \cv(w)P(z) :: \: - \: \frac{\a \, wz}{w^2 - z^2}
\; C \, \cv(w) \qquad \mid w \mid > \mid z \mid \er
We expand $\cv(z)$ in  modes as 
\be \cv(z) = \sum_{n \in { \IZ}  } \; z^{-n} A_n \qquad
\qquad
 A_n = \oint \frac{dz \; z^n}{2 \pi i \, z}\: \cv(z)
 \lab{modos}\ee
That means that $b_{\mathbf{M}} A_n $ and $A_n b_{\mathbf{M}} $
can be written as the doble integral of the same term
\br b_{\mathbf{M}}\; A_n &=&  - \left({\oint \frac{d\zeta \;
\zeta^n}{2 \pi i \, \zeta} \oint  \frac{dz \; z^{\mathbf{M}}}{2
\pi i \, z}}\right)_{\mid z \mid > \mid \zeta \mid} \: \left\{ ::
P(z)\cv(\zeta) :: \: + \: \frac{\a \; \zeta z}{z^2 - \zeta ^2} \:
C  \, \cv(\zeta) \right\}
\nonu\\
 A_n \; b_{\mathbf{M}}  &=&  - \left({\oint \frac{d\zeta \; \zeta^n}{2
\pi i \, \zeta} \oint  \frac{dz \; z^{\mathbf{M}}}{2 \pi i \,
z}}\right)_{\mid \zeta \mid > \mid z \mid} \: \left\{ ::
P(z)\cv(\zeta) :: \: + \: \frac{\a \; \zeta z}{z^2 - \zeta ^2} \:
C  \, \cv(\zeta) \right\} \er
where we use
\be \oint \frac{dz}{2 \pi i \, z} \; z^{\mathbf{M}} \; P(z) = -
b_{\mathbf{M}} \qquad {\mathbf{M}} \:\: {\rm odd} \lab{pz} \ee
Notice that
\be \left({\oint \frac{d\zeta \; \zeta^n}{2 \pi i \,
\zeta} \oint \frac{dz \; z^{\mathbf{M}}}{2 \pi i \,
z}}\right)_{\mid z \mid > \mid \zeta \mid}  - \left({\oint
\frac{d\zeta \; \zeta^n}{2 \pi i \, \zeta} \oint  \frac{dz \;
z^{\mathbf{M}}}{2 \pi i \, z}}\right)_{\mid \zeta \mid > \mid z
\mid} = \ee
\begin{center}  \begin{picture}(300,100)(0,0)
\Line(0,0)(0,100) \ArrowLine(0,90)(0,100)
\Line(-50,50)(50,50)\ArrowLine(40,50)(50,50) \CArc(0,50)(40,0,360)
\Text(15,70)[]{$\bullet$} \Text(7,75)[]{$\zeta$}
\Text(-25,35)[]{-$\zeta$} \Text(-15,30)[]{$\bullet$}
\ArrowArc(0,50)(40,100,140) \Text(65,50)[]{$-$}
\Line(120,0)(120,100) \ArrowLine(120,90)(120,100)
\Line(80,50)(160,50)\ArrowLine(150,50)(160,50)
\CArc(120,50)(20,0,360) \Text(140,80)[]{$\bullet$}
\Text(130,85)[]{$\zeta$} \Text(90,25)[]{-$\zeta$}
\Text(100,20)[]{$\bullet$} \Text(180,50)[]{$=$}
 \ArrowArc(120,50)(20,100,140)
\Line(265,0)(265,100) \ArrowLine(265,90)(265,100)
\Line(200,50)(330,50) \ArrowLine(320,50)(330,50)
\Text(230,20)[]{$\bullet$} \Text(220,25)[]{-$\zeta$}
\CArc(230,20)(18,0,360) \ArrowArc(230,20)(18,100,140)
\Text(290,75)[]{$\zeta$} \Text(300,70)[]{$\bullet$}
\CArc(300,70)(18,0,360)\ArrowArc(300,70)(18,100,140)
\end{picture}   \end{center}
The only contribution comes from the poles on $z = \pm \zeta$; we
eliminate the poles on $z =0$. So we have
 \be \lb b_{\mathbf{M}}
, \; A_n   \rb = - {\oint \frac{d\zeta \; \zeta^n}{2 \pi i \,
\zeta} \:\: \:\: \oint \frac{dz \; z^{\mathbf{M}}}{2 \pi i \,
z}}  \:\: \:\:  \left\{ :: P(z)\cv(\zeta) :: \: + \: \frac{\a \;
\zeta z}{z^2 - \zeta ^2} \: C  \, \cv(\zeta) \right\}\ee
The first term in r.h.s. is normal ordered, what guarantees that
there is no contribution of this part on $z = \zeta$. The second
term  has a simple pole on $z = \pm \zeta$. So, by the residue
theorem we have
\br \lb b_{\mathbf{M}} , \; A_n   \rb &=& - \frac{\a}{2}\, C\,
\oint \frac{d\zeta \; \zeta^n}{2 \pi i \, \zeta} \:
\zeta^{\mathbf{M}} \, \cv(\zeta) \: + \: \frac{\a}{2}\, C\, \oint
\frac{d\zeta \; \zeta ^n}{2 \pi i \, \zeta } \: (- \zeta
)^{\mathbf{M}} \, \cv(\zeta
) \nonu\\
&=& - \a \, C \, A_{n + {\mathbf{M}}} \er
once ${\mathbf{M}}$ is odd and we used \rf{modos}, \rf{pz}. This
relation corresponds to \rf{comu1}, \rf{comu2}, but in order to
reproduce the algebra, we must impose
 \be 
\a \, C =  2 \lab{alfa}
 \quad \Rightarrow \quad
\lb b_{\mathbf{M}} , \;
A_n   \rb = - 2 \,  A_{n + {\mathbf{M}}}
\lab{comrelban}
 \ee
Now we must verify the remaining commutation relations, namely
$\lb A_m, A_n \rb$. To this, we must evaluate 
\be \cv(z)\cv(\zeta) = \: :: \cv(z)\cv(\zeta) :: \: e^{\frac{\a^2}{2}\, C
\; \ln{(\frac{z-\zeta}{z+\zeta})}  }  \quad \mid z \mid > \mid
\zeta \mid \ee
from where we conclude that
\br \lb A_m , \, A_n \rb &=& \oint \frac{dz \; z^m}{2 \pi i \, z}
\; \:\: \oint \frac{d\zeta \; \zeta^n}{2 \pi i \, \zeta}
\;   \:\: \: :: \cv(z)\cv(\zeta) :: \times \nonu\\
& & \times \:\: \left[ \left. \left( \frac{z-\zeta}{z+\zeta}
\right)^{\frac{\a^2}{2} \, C} \right|_{\mid z \mid
> \mid \zeta \mid} - \left. \left( \frac{\zeta - z}{\zeta + z}
\right) ^{\frac{\a^2}{2} \, C} \right| _{\mid \zeta \mid > \mid z
\mid } \right] \er
Since we need double poles, we set $\frac{\a ^2}{2} \, C = 2$. But from
\rf{alfa} we have $\a C = 2$, and so we get that $\a =2$ and $C=1$. Then 
\be \lb A_m , \, A_n \rb = \oint \frac{dz \; z^m}{2 \pi i \, z}
\;   \:\: \oint \frac{d\zeta \; \zeta^n}{2 \pi i \, \zeta}
\;   \:\: \: :: \cv(z)\cv(\zeta) :: \: \left( \frac{z -
\zeta}{z + \zeta} \right)^2 \ee where we take the contour to be

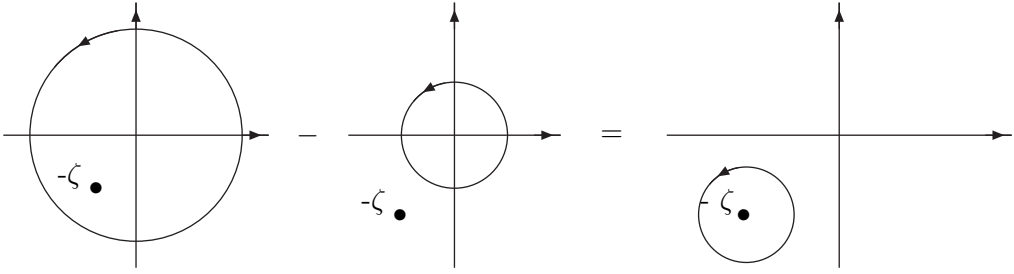
\begin{center}  \begin{picture}(300,100)(0,0)
\Line(0,0)(0,100) \ArrowLine(0,90)(0,100)
\Line(-50,50)(50,50)\ArrowLine(40,50)(50,50) \CArc(0,50)(40,0,360)
\Text(-25,35)[]{-$\zeta$} \Text(-15,30)[]{$\bullet$}
\ArrowArc(0,50)(40,100,140) \Text(65,50)[]{$-$}
\Line(120,0)(120,100) \ArrowLine(120,90)(120,100)
\Line(80,50)(160,50)\ArrowLine(150,50)(160,50)
\CArc(120,50)(20,0,360)  \Text(90,25)[]{-$\zeta$}
\Text(100,20)[]{$\bullet$} \Text(180,50)[]{$=$}
 \ArrowArc(120,50)(20,100,140)
\Line(265,0)(265,100) \ArrowLine(265,90)(265,100)
\Line(200,50)(330,50) \ArrowLine(320,50)(330,50)
\Text(230,20)[]{$\bullet$} \Text(220,25)[]{- $\zeta$}
\CArc(230,20)(18,0,360) \ArrowArc(230,20)(18,100,140)
\end{picture}   \end{center}

According with the Cauchy's integral formula 
\be 
f'(z_0) =
\frac{1}{2 \pi i } \oint \frac{f(z)}{(z - z_0)^2} \: dz
\ee 
what
implies 
\br 
\lb A_m , \, A_n \rb &=&  \oint \frac{d\zeta \;
\zeta^n}{2 \pi i \, \zeta} \;  \zeta ^{n + m } \:\: 4 (-1)^m\left[
m -  \a \, \, P(\zeta) \right] 
\er
where we have used the fact that $\: :: \cv(-\zeta)\cv(\zeta) ::\:=1$. 
Using  \rf{pz} one gets
\br
\lb A_{2m} , \, A_{2n} \rb &=& 8 m \d_{m+n,0}\nonu\\
\lb A_{2m+1} , \, A_{2n+1} \rb &=& -4\( 2m+1\) \d_{m+n+1,0}\nonu\\
\lb A_{2m+1} , \, A_{2n} \rb &=& -8 \, b_{2\(m+n+\)+1}
\lab{comrelan}
\er

Therefore, from \rf{comrelban} and \rf{comrelan}, we see that $\pm
A_n/2$ satisfy the same algebra as $F_n$, given in
\rf{comu1}-\rf{hei2}. So we
have reproduced the commutation relations for the $sl(2)$ Kac-Moody
algebra for $C=1$.  
The arbitrariness in the sign of $A_n$ is related to the normalization
of the expectation values of the 
operators on the highest weight states. Since the operator $\cv\(z\)$
introduced in \rf{vertexdef} is normal ordered, it satisfy
\be
\bra{\l} \cv\(z\) \ket{\l} = 1 
\ee
Then, in order to reproduce \rf{expectvaluev} we must normalize $\cv\(z\)$
differently in different highest weight representations. In fact, we have to
make the identification  
\be
V\(z\) = c_{\l} \cv\(z\)
\lab{normalv}
\ee
where
\be
c_{\l_0} = 1 \qquad \qquad c_{\l_1} = -1
\lab{clpm}
\ee

\subsection{Vertex Operator properties}
\label{app:com}

Using \rf{qpm}

\be Q_{>}(z) = \sum_{n = 0}^{\infty} \: \frac{z^{-
\mathbf{N}}}{\mathbf{N}}\: b_{\mathbf{N}} \qquad \qquad
Q_{<}(z) =
- \sum_{n = 0}^{\infty} \:
  \frac{z^{
\mathbf{N}}}{\mathbf{N}} \: b_{- \mathbf{N}}  \qquad \qquad
{\mathbf{N}} = 2n+1 \ee

\br
 \cv (z) \cv(w) &=& e^{2 \, Q _{<}(z) } \, e^{2 \, Q _{>}(z) } \:\:
e^{2\, Q _{<}(w) } \, e^{2 \, Q _{>}(w) } \nonu\\
&=&
  e^{2 \, Q _{<}(z) } \quad e^{2 \, Q _{>}(z) } \:
e^{2\, Q _{<}(w) } \, e^{- 2 \, Q _{>}(z) } \quad e^{2 \,
Q_{>}(z)}\, e^{2 \, Q _{>}(w) }\lab{2v}\er

\br e^{2 \, Q _{>}(z) } \: e^{2\, Q _{<}(w) } \, e^{- 2 \,
Q_{>}(z) } &=& e^{ e^{2 \, Q _{>}(z) } \: 2\, Q _{<}(w) \:
e^{-2\, Q _{>}(z) } } \nonu\\
 &=& e^{   2\, Q _{<}(w) + \lb 2 \, Q _{>}(z) , \, 2 \,  Q _{<}(w)  \rb }
 \nonu\\
 &=& e^{   2 \, Q _{<}(w)} \;  e^{ - 2 \, \ln \left( \frac{z + w}{z-w} \right)
   } 
\lab{flip1}
  \er
where we used
\br \lb 2 \, Q _{>}(z) , \, 2 \,  Q _{<}(w)  \rb  &=& - 4
\sum_{n,m \geq 0} ^{\infty} \; \frac{z^{-
\mathbf{N}}}{\mathbf{N}} \; \frac{w^{ \mathbf{M}}}{\mathbf{M}} \:
\lb b_{\mathbf{N}} , \,
b_{- \mathbf{M}}\rb \nonu\\
&=& - 4 \sum_{n,m \geq 0} ^{\infty} \;
\frac{z^{-\mathbf{N}}}{\mathbf{N}} \;
\frac{w^{\mathbf{M}}}{\mathbf{M}} \:
{\mathbf{N}} \; \d _{ {\mathbf{N}}, \, {\mathbf{M}} }  \nonu\\
&=& - 4  \sum_{n \geq 0} ^{\infty} \; \frac{1}{\mathbf{N}} \left(
\frac{w}{z} \right) ^{\mathbf{N}} \nonu\\
&=& - 2 \, \ln \left( \frac{z + w}{z-w} \right)
 \er
Then, from \rf{flip1} we have 
\be
e^{2 \, Q _{>}(z) } \: e^{2\, Q _{<}(w) } = 
e^{   2 \, Q _{<}(w)} \;  \, e^{ 2 \,Q_{>}(z) }
e^{ - 2 \, \ln \left( \frac{z + w}{z-w}\right)} 
\lab{flip2}
\ee

So, returning to \rf{2v} and using \rf{flip2}
 \br  
\cv (z) \cv(w) &=& e^{2 \, Q _{<}(z) } \;
e^{2\, Q _{<}(w) } \:\:\: e^{2 \, Q_{>}(z)} \; e^{2 \, Q _{>}(w) }
\;\:\:\:  e^{ - 2 \, \ln \left( \frac{z + w}{z-w} \right) }
\nonu\\ &=& \: ::\cv (z) \cv(w) :: \:\:\: \left( \frac{z - w}{z+w}
\right)^2 
\lab{vzvw}
\er
By the same argument,
\br
  \cv (z) \cv(w) \cv(k)=  \: ::\cv (z) \cv(w)\cv(k) :: \:\:\: \left( \frac{z -
 w}{z+w} 
\right)^2 \left( \frac{z - k}{z+k} \right)^2 \left( \frac{w
-k}{w+k} \right)^2
\er
and so on. Therefore, from \rf{normalv} we get that
\be 
\bra{\l} \;  V(\mu_1) \dots V(\mu_n) \;
\ket{\l} = c_\l ^n \:\:  \; \prod_{j>i=1}^n\, 
\( \frac{\mu_i - \mu_j}{\mu_i + \mu_j} \)^{2}  
\lab{nvexpectvalue}
\ee
where we have used that fact that
\be
 \bra{\l} \; ::   \cv(\mu_1) \dots  \cv(\mu_n) :: \; \ket{\l} = 1
\ee

\section{The numerical check of equations \rf{nldeq}}
\label{sec:nume}

As we have seen in section \ref{sec:check},  the existence of the exact
solution relies on the validity of the non-linear differential
equations \rf{nldeq} satisfied by the quantities $I_{2n+1}$. We could
not find in the literature any reference to those equations, and 
we were not able to prove them analitically. Therefore, we present here a
numerical check of the first three equations in \rf{nldeq}, at some
specific values of the argument of those functions. 

In order to perform the numerical calculation it is better to make a change of
integration variables in the expressions for $I_N$ and $J_N$. We define
\be
\phi_i = \ln \frac{1+x_i}{1-x_i} \qquad \qquad i=1,2, \ldots N-1
\ee
Then
\br
x_i = -1 &\ra & \phi_i = - \infty\nonu\\
x_i = 1 &\ra & \phi_i =  \infty
\er
We have that
\be
d \phi_i = \frac{2}{1-x_i^2} dx_i
\ee
In addition we have
\br
\cosh \( \h \phi_i\) &=& \frac{1}{\sqrt{1-x_i^2}}\nonu\\
\cosh \( \h \sum_{i=1}^{N-1}\phi_i\) &=& 
\h \; \frac{\prod_{i=1}^{N-1}\(1+x_i\) + \prod_{i=1}^{N-1}\(1-x_i\)}
{\prod_{i=1}^{N-1}\sqrt{1-x_i^2}}\nonu\\
\cosh \(  \sum_{i=j_1}^{j_2}\phi_i\) &=& 
\h \( \prod_{i=j_1}^{j_2}\frac{\(1+x_i\)}{\(1-x_i\)} 
+ \prod_{i=j_1}^{j_2}\frac{\(1-x_i\)}{\(1+x_i\)} \)
\nonu\\
&=& \h \frac{\(\prod_{i=j_1}^{j_2}\(1+x_i\)\)^2 + 
\(\prod_{i=j_1}^{j_2}\(1-x_i\)\)^2}
{\prod_{i=j_1}^{j_2}\(1-x_i^2\)}
\er
We then get that $I_N$ and $J_N$ given in \rf{finalin} and \rf{jndef}
respectively, become 
\br
I_N \(z\) &=& \int_{-1}^{1} d^{N-1}x_i \; 
\frac{K_0\(z\sqrt{w_N}\)}{v_N}\nonu\\
J_N \(z\) &=& \int_{-1}^{1} d^{N-1}x_i \; 
\frac{w_N\; K_0\(z\sqrt{w_N}\)}{v_N}
\er
where
\be
w_N = N + \sum_{l=0}^{N-2} \sum_{j=1}^{N-l-1} \( 
\prod_{i=j}^{j+l} \frac{\(1+x_i\)}{\(1-x_i\)} + 
\prod_{i=j}^{j+l} \frac{\(1-x_i\)}{\(1+x_i\)}\)
\ee
and
\be
v_N = \h \; \( \prod_{i=1}^{N-1} \( 1+x_i\) +\prod_{i=1}^{N-1} \( 1-x_i\)\) 
\ee

In order to evaluate the multidimensional integrals we used the Monte Carlo 
method. The program was written in Mathematica and did not make use of the
command {\tt NIntegrate}, since we found some limitations on it regarding 
the  number of iterations.  We found it better to write a routine to do it
directly, using Mathematica random number generator and its package
for Bessel functions.   The functions $I_N$
and $J_N$ are defined as $(N-1)$-dimensional integrals. We have
evaluated them for $N=3,5, \, {\rm and}\, 7$,  by sampling the 
integration region  with $10^6$ points, irrespective of the 
dimensionality  of the integrals. We evaluated those functions for the
arguments $z=0.001, 0.01, 0.1, 0.5, 1, 5, 10, \, {\rm and}\, 100$. 
In order to estimate the errors in
the calculations, we evaluated each integral for each value of the
argument, $11$ times with the $10^6$ point sampling. We then
calculated the average (${\bar v} = \sum_{i=1}^n v_i/n$) and the
standard deviation  
($\sigma^2 = \sum_{i=1}^n \( v_i - {\bar v}\)^2 /\(n-1\)$). 

Using \rf{niceinjnrel} one can write the equations \rf{nldeq} as 
\be
J_{2n+1} = L_{2n+1}
\lab{eqtocheck} 
\ee
where 
\br
L_3&\equiv& I_3 + 8 I_1^3 \nonu\\
L_5&\equiv& I_5 + \frac{40}{3} I_1^2 \, I_3 + \frac{32}{3} I_1^5\nonu\\
L_7&\equiv& I_7 + \frac{224}{9} I_1^4 \, I_3 + 
\frac{56}{9} I_1 \, I_3^2 + \frac{56}{5} I_1^2 \, I_5 + 
\frac{256}{45} I_1^7
\er
Therefore, we check the relation \rf{eqtocheck} numerically. The
errors in $L_{2n+1}$ are given by 
\br
\Delta L_3&=& \Delta I_3  \nonu\\
\Delta L_5&\equiv& \Delta I_5 + \frac{40}{3} I_1^2 \, \Delta I_3 \nonu\\
\Delta L_7&\equiv& \Delta I_7 + \frac{224}{9} I_1^4 \, \Delta I_3 + 
\frac{112}{9} I_1 \, I_3\, \Delta I_3 + \frac{56}{5} I_1^2 \, \Delta I_5 
\er
Since $I_1 = K_0$, and since it is evaluated with the Mathematica
package for Bessel function, its error was taken to be negligible as 
 compared to the other ones. Its values, for the arguments used in
 the simulations are given in table \ref{tab:i1}.
In table \ref{tab:i3i5i7} we show the results of the numerical
calculations of the integrals $I_3$, $I_5$ and $I_7$, with the
corresponding standard deviations. 

The check of the relations \rf{eqtocheck}, or equivalently \rf{nldeq},
is given in tables \ref{tab:j3l3}, \ref{tab:j5l5} and
\ref{tab:j7l7}. Notice that the agreement is quite good. Except 
for the values of $J_5$ and $L_5$ at $z=0.001$, which differ by $1.45$
standard deviations, all the other values agree well inside one
standard deviation. So, our numerical calculations strongly indicate
that the relations \rf{nldeq} should hold true. It would be very
interesting to find an analytical proof of those equations.

\begin{table} 
\begin{center}
\begin{tabular}{|c|c|}
\hline
$z$ & $I_1 \equiv K_0$ \\
\hline
$0.001$ & $7.02369$ \\
$0.01$ & $4.72124$ \\
$0.1$ & $2.42707$ \\
$ 0.5$ & $0.924419$ \\
$1$ & $0.421024$ \\
$ 5$ & $3.6911 \; 10^{-3}$ \\
$10$ & $1.77801 \; 10^{-5}$ \\
$100$ & $4.65663\;{10}^{-45}$ \\
\hline
\end{tabular}
\end{center}
\caption[]{\parbox[t]{.7\textwidth}{The numerical values of $I_1$ for
        some values of the argument}}
        \protect\label{tab:i1}
\end{table}

\begin{table} 
\begin{center}
\begin{tabular}{|c|c|c|c|}
\hline
$z$ &  $I_3$ & $I_5$ & $I_7$\\
\hline
$0.001$ & $26.237 \pm 0.026$ & $155.52 \pm 0.44$ & $1027.0 \pm 3.2$ \\
$0.01$ & $14.92 \pm 0.01$ & $73.30 \pm 0.12$ & $390.4 \pm 1.0$ \\
$0.1$ & $4.455 \pm 0.002$ & $11.429 \pm 0.011$ & $30.118 \pm 0.035$ \\
$ 0.5$ & $0.43063 \pm 0.00044$ & $0.24327 \pm 0.00024$ & $0.13817\pm 0.00031$\\
$1$ & $0.049517 \pm 0.000035$ & $0.006656 \pm 0.000010$ &
$\( 8.974 \pm 0.028 \) 10^{-4}$ \\
$ 5$ & $\( 4.4348 \pm 0.0068\) 10^{-8}$  & $\( 5.548 \pm 0.018 \) 10^{-13}$ &
$\( 6.9 \pm 0.1\) 10^{-18}$ \\
$10$ & $\(5.25  \pm  0.02\) 10^{-15}$ & $\(1.589 \pm 0.017\) 10^{-24}$ &
$\(4.79 \pm 0.21\) 10^{-34}$ \\
$100$ &  $\( 9.990 \pm 0.065\)  10^{-134}$  &
$\( 2.21  \pm  0.27\)  10^{-222}$    &
$\(3.8 \pm 2.8\)  10^{-311}$    \\
\hline
\end{tabular}
\end{center}
\caption[]{\parbox[t]{.7\textwidth}{The numerical values of $I_3$,
        $I_5$ and $I_7$ with corresponding standard deviations}}
        \protect\label{tab:i3i5i7}
\end{table}

\begin{table} 
\begin{center}
\begin{tabular}{|c|c|c|}
\hline
$z$ &  $J_3$ & $L_3$ \\
\hline
$0.001$ & $2783. \pm 338.$ & $2798.19 \pm 0.03$\\
$0.01$ & $863.9 \pm 27.8$ & $856.82 \pm 0.01 $\\
$0.1$ & $118.83 \pm 0.27$ & $118.831 \pm 0.002$\\
$ 0.5$ & $6.7521 \pm 0.0074$ & $6.75033 \pm 0.00044$\\
$1$ & $0.64645 \pm 0.00040$ & $0.646569 \pm 0.000035$ \\
$ 5$ &$\(4.4631 \pm 0.0063\) 10^{-7}$ &  
$\(4.46654 10   \pm 0.00068 \) 10^{-7}$\\
$10$ & $\(5.022   \pm 0.021\) 10^{-14}$ & 
$\(5.0220 \pm 0.0022\) 10^{-14}$ \\
$100$ & $\(9.051 \pm 0.059\) 10^{-133}$  & 
$\(9.0770 \pm 0.0065\) 10^{-133}$  \\
\hline
\end{tabular}
\end{center}
\caption[]{\parbox[t]{.7\textwidth}{Comparison of the numerical values
    of $J_3$, and $L_3$ with cor\-res\-pon\-ding standard deviations}}
        \protect\label{tab:j3l3}
\end{table}

\begin{table} 
\begin{center}
\begin{tabular}{|c|c|c|}
\hline
$z$ &  $J_5$ & $L_5$ \\
\hline
$0.001$ & $181810. \pm 12383.$ & $199742. \pm 18.$ \\
$0.01$ & $29616. \pm 318.$ &  $29529.2 \pm 3.1$ \\
$0.1$ & $1260.3 \pm 1.7$ &  $1259.64 \pm 0.17$ \\
$ 0.5$ & $12.35 \pm 0.01$ &   $12.3505 \pm 0.0052$ \\
$1$ & $0.26474 \pm 0.00032$ &  $0.264801 \pm 0.000093$ \\
$ 5$ & $\(1.5883 \pm  0.0052\) 10^{-11}$ & $\(1.5919 \pm 0.0014\) 10^{-11}$ \\
$10$ &$\(4.273 \pm  0.046\) 10^{-23}$ & $\(4.268 \pm  0.011\) 10^{-23}$ \\
$100$ & $\(5.56  \pm 0.67\) 10^{-221}$ &
$\(5.45 \pm 0.46\) 10^{-221}$ \\
\hline
\end{tabular}
\end{center}
\caption[]{\parbox[t]{.7\textwidth}{Comparison of the numerical values
    of $J_5$, and $L_5$ with cor\-res\-pon\-ding standard deviations}}
        \protect\label{tab:j5l5}
\end{table}

\begin{table} 
\begin{center}
\begin{tabular}{|c|c|c|}
\hline
$z$ &  $J_7$ & $L_7$ \\
\hline
$0.001$ & $\(6.40 \pm .65\) 10^{6}$ &  $\(6.5034 \pm 0.0019\) 10^{6}$\\
$0.01$ & $505613. \pm 4987.$ & $507197. + 166.$\\
$0.1$ & $7753. \pm 11.$ &  $7753.3 \pm 2.9$ \\
$ 0.5$ & $14.641 \pm 0.027$ &  $14.642 \pm 0.013$\\
$1$ & $0.07265 \pm 0.00018$ & $0.072600 \pm 0.000059$ \\
$ 5$ &  $\(3.932  \pm 0.053\) 10^{-16}$ & $\(3.9474 \pm 0.0083\)10^{-16}$ \\
$10$ &  $\(2.54  \pm 0.11\) 10^{-32}$ & $\(2.542 \pm 0.016\) 10^{-32}$\\
$100$ & $\(1.9 \pm  1.4\) 10^{-309}$ &
$\(2.3 \pm 1.0\) 10^{-309}$ \\
\hline
\end{tabular}
\end{center}
\caption[]{\parbox[t]{.7\textwidth}{Comparison of the numerical values
    of $J_7$, and $L_7$ with cor\-res\-pon\-ding standard deviations}}
        \protect\label{tab:j7l7}
\end{table}

\end{document}